\documentclass[aps,prc,twocolumn,superscriptaddress,preprintnumbers,amsmath,amssymb,showkeys,floatfix,nofootinbib,reprint]{revtex4-1}

\usepackage{soul}
\usepackage{amsfonts}
\usepackage{amssymb}
\usepackage{graphics}
\usepackage{epstopdf}
\usepackage{longtable}
\usepackage{epsfig}
\usepackage[normalem]{ulem}
\usepackage {dcolumn}
\usepackage {graphicx}
\usepackage {dcolumn}
\usepackage {bbm}
\usepackage [version = 4] {mhchem}
\usepackage {color}
\usepackage {xcolor}

\def\bcbaux#1#2 #3\endbcb{%
  \colorbox{#1}{\strut#2}%
  \ifx\relax#3\relax\def\next{}\else%
    \colorbox{#1}{\strut}%
   \allowbreak%
   \def\next{\bcbaux{#1}#3\endbcb}%
  \fi%
  \next%
}
\usepackage {multirow}
\usepackage {textcomp}
\newcommand {\bs} {\boldsymbol}

\usepackage[most]{tcolorbox}
\newtcolorbox{highlighted}{colback=yellow,coltext=red,breakable}

\begin {document}

\title {Isomer production by multi-photon excitation}

\author {Weitao Liu}
\affiliation {School of Physics, Nankai University, Tianjin 300071, China}

\author {Yuanbin Wu}
\email{yuanbin@nankai.edu.cn}
\affiliation {School of Physics, Nankai University, Tianjin 300071, China}

\date{\today}

\begin {abstract}

	The multi-photon excitation to the $8$-eV nuclear isomeric state $^{229\text{m}}$Th in the direct laser-nucleus interaction is investigated theoretically. We solve the time-dependent Schr\"odinger equation with the method which allows us to study the $n$-photon absorption in the nuclear excitation in the direct laser-nucleus interaction. Based on the laser facilities available currently or in the near future, we analyze the impact of the laser parameters on the excitation probability of the multi-photon excitation. The possibilities of the $2$-, $3$- and $4$-photon excitations to the isomeric state $^{229\text{m}}$Th from the ground state are discussed in details. Our results show the strong impact of the laser intensity and pulse duration on the multi-photon excitation probability. The onset of high-order effects in the multi-photon excitation in the direct laser-nucleus interaction is also revealed. Our findings open new possibilities to study the multi-photon laser-nucleus interaction in high-power laser facilities.

\end {abstract}

\maketitle

\section {Introduction}

	The concept of the multi-photon excitation, firstly described by G\"oppert-Mayer \cite {GoeppertMayer1931}, provides an alternative method to manipulate quantum states other than the direct $1$-photon excitation \cite{Dirac243}. With the development of the laser technology in the past decades, theoretical and experimental investigations of the multi-photon absorption/excitation in atomic, molecular, or condensed matter systems \cite{PhysRevLett.7.229, PhysRevLett.9.453,PhysRevA.13.1817, PhysRevA.13.1829,PhysRev.134.A499,PhysRevB.40.1403, 7400910} have attracted a great deal of attention. Moreover, applications of the multi-photon absorption have also been proposed in microscopy \cite{Larson2011, Horton2013}, photodynamic therapy \cite{C6CS00442C, MCKENZIE20192, adma.201701076}, optical data storage \cite{Cumpston1999, C7TC00582B} and laser frequency conversion \cite{Thielking2023, BERDAH1996118, 1068619}. In contrast, for nuclear systems, discussions on the multi-photon excitation are so far still limited, although it has been shown that the concept has interesting features \cite{PhysRevLett.42.1397,PhysRevC.20.1942,PhysRevC.23.50,yang2024new,item_3026914,PhysRevLett.133.152503,lu2025nuclearexcitationcontrolinduced}. One of the main reasons is that, due to the difference between the nuclear energy levels and the conventional laser technology \cite{Wense2020a}, the narrow-band direct laser-nucleus excitation used to be a great challenge in experiments. Another reason is that, due to the small matrix elements of multipole operators, the multi-photon absorption in the laser-nucleus excitation is difficult. The $8$-eV nuclear isomeric state $^{229\text{m}}$Th offers an excellent candidate for this topic \cite{PhysRevLett.133.152503}.

	$^{229\text{m}}$Th has attracted a great deal of attention in recent decades. With the lowest known nuclear excitation energy at about $8\ \text{eV}$ and the long lifetime, it is considered to be an excellent candidate for a nuclear clock \cite{Wense2020, Kazakov2012, Thirolf_2019, Beeks2021, Peik2021}. This isomeric state, due to its extremely low excitation energy, provides opportunities for the direct laser manipulation of nuclear states. In recent experiments \cite {PhysRevLett.132.182501, PhysRevLett.133.013201, Zhang2024}, the direct $1$-photon excitation to the isomeric state from the ground state has been realized. Moreover, by exploiting the nuclear hyperfine mixing effects, the isomeric excitation has been shown theoretically to have highly nonlinear effects and high probability in the interaction of an intense laser with hydrogen-like $^{229}$Th ions \cite{PhysRevLett.133.152503}. This offers an excellent case for the studies of the multi-photon excitation in nuclear systems by utilizing the interplays between the nuclear and atomic systems.

	The multi-photon excitation in the direct laser-nucleus interaction without the involvement of the atomic degree of freedom is still a great challenge and requires further investigations. Tuning the photon energy to $1/n$ of the excitation energy ($n$ is an integer), $^{229\text{m}}$Th might be produced by the $n$-photon excitation process. The process is sometimes named as degenerate multi-photon excitation \cite{Rumi2010,He2002}, when $n>1$. It is a high-order process which is expected to be small in the quantum perturbation theory \cite{Zelevinsky2010}. However, intense and short laser pulses with an intensity on the level of $10^{23}\ \text{W}/\text{cm}^2$ can be achieved experimentally \cite{Yoon:21}. Higher-powered petawatt and exawatt lasers in the plan of many countries \cite{Danson2015, Danson2019, RevModPhys.94.045001} may provide laser beams with higher intensities in the future. Furthermore, another kind of laser facilities, such as Laser MegaJoule (LMJ) \cite{MIQUEL2020, CEA/DAM2020, RevModPhys.94.045001} and National Ignition Facility (NIF) \cite{Hogan_2001, Haynam2007, RevModPhys.94.045001, Cerjan2018}, can provide laser beams at lower intensities with the full width at half minimum (FWHM) of the pulse duration as long as several tens of nanoseconds, which may also play important roles in the nuclear multi-photon excitation. As a consequence, the degenerate multi-photon excitation may provide a feasible routine to produce considerable $^{229\text{m}}$Th yield.

    In this article, we study theoretically the multi-photon excitation to the isomeric state $^{229\text{m}}$Th from the ground state in the direct laser-nucleus interaction, by solving the time-dependent Schr\"odinger equation (TDSE). The TDSE is solved in the present work with a method which allows us to study the $n$-photon absorption in the nuclear excitations in the direct laser-nucleus interaction. Based on the laser facilities available currently or in the near future, we analyze in details the impact of the laser parameters, such as the laser intensity and the laser pulse duration, on the probability of the multi-photon nuclear excitation. This article is organized as follows: in Sec. \ref {sec:theoretical_framework}, we introduce our theoretical approach for the nuclear multi-photon excitation in the direct laser-nucleus interaction; in Sec. \ref {sec:numerical_results}, we demonstrate our numerical results, which include the general multi-photon excitation processes and the possible yields; in Sec. \ref{sec:conclusion}, we briefly summarize our work. Useful information on the direct laser-nucleus interaction is presented in Appendix \ref{sec:laser-nucleus interaction}.

\section {Theoretical approach}
	\label {sec:theoretical_framework}

    In this section, we describe our theoretical approach for the nuclear multi-photon excitation in the direct laser-nucleus interaction. We note that, when the photon energy is set to $1/n$ of the excitation energy, every path with the net photon absorption number $n$ contributes to the excitation probability. The order-by-order analysis requires us to find out and sum over all these paths in the $N$-th order, with $N-n$ a positive even number. Herein and after, the phrase \emph{$n$-photon absorption} refers to the process with the net photon absorption number $n$. 

    The theoretical approach in the present work is based on the TDSE. Numerical solutions to the TDSE \cite{PhysRevE.84.056703} have been widely used in studies of laser-atom interactions \cite{PhysRevA.95.023401,Tong2011,Zakavi2024}. In this article, we improve the numerical method by sorting the processes by the photon absorption number. This makes the contributions of various processes distinguishable. 

	\subsection {The time-dependent Schr\"odinger equation}
	
		To describe a nucleus subjected to a Gaussian-shaped laser pulse with the frequency $\omega$, we use the TDSE
		\begin {equation}
			\begin {aligned}
				& i\frac{\partial}{\partial{t}}|\psi(t)\rangle = H_0 |\psi(t)\rangle \\
				& \qquad + \frac{ie^{-t^2/\sigma^2}}{2} \left(e^{-i\omega{t}}U^+-e^{i\omega{t}}U^-\right)|\psi(t)\rangle,
			\end {aligned}
            \label {eq:Schroedinger}
		\end {equation}
		where $H_0$ is the Hamiltonian of the free nucleus, which determines the eigenstates of the free nucleus. And $\sigma = \Gamma_{\text{I}} / (\sqrt{2\ln2})$, with $\Gamma_{\text{I}}$ the FWHM of the Gaussian-shaped intensity peak in the time domain. $U^+$ and $U^-$ are Hermitian conjugate operators describing the photon absorption and emission, respectively (their explicit forms are derived in Appendix \ref{sec:laser-nucleus interaction}). With the expansion $|\psi(t)\rangle = \sum_i c_i(t) e^{-iE_it} |\phi_i\rangle$, where $E_i$ and $|\phi_i\rangle$ are the eigenvalue and the eigenvector of $H_0$, the equation has the formal solution
		\begin {equation}
			\begin {aligned}
				c_{k}(t) =\ & \frac{1}{2} \int_{t_0}^t d\tilde{t} e^{-{\frac{\tilde{t}^2}{\sigma^2}}} \left(e^{-i\omega\tilde{t}}U^+_{kj}-e^{i\omega\tilde{t}}U^-_{kj}\right) \\
				& \qquad \times e^{iE_k\tilde{t}} c_{j}(\tilde{t}) e^{-iE_j\tilde{t}}.
			\end {aligned}
         			\label {eq:formal_solution}
		\end {equation}
        
		Since the coefficient $c_k(t)$ appears in both two sides, Eq. \eqref{eq:formal_solution} cannot be immediately used to obtain the numerical result. A time shift $t\rightarrow{t}+\delta{t}$ leads to 
		\begin {equation}
			\begin {aligned}
				c_{k}(t+\delta{t}) &= \frac{1}{2} \left(\int_{t}^{t+\delta{t}}+\int_{t_0}^{t}\right) d\tilde{t} e^{-{\frac{\tilde{t}^2}{\sigma^2}}} \\
				& \times \left(e^{-i\omega\tilde{t}}U^+-e^{i\omega\tilde{t}}U^-\right) e^{-i\left(E_j-E_k\right)\tilde{t}} c_{j}(\tilde{t}) \\
				& = \frac{\delta{t}}{2} e^{-{\frac{t^2}{\sigma^2}}} \left[e^{-i(E_j+\omega-E_k){t}}U^+_{kj}\right. \\
				& \left. -e^{-i(E_j-\omega-E_k){t}}U_{kj}^-\right] c_{j}(t)  \\
				& + c_k(t) + \mathcal{O} [(\Delta{E}\delta{t})^2],
			\end {aligned}
			\label {eq:time_shift}
		\end {equation}
		where $\Delta{E}$ is the difference between the highest and the lowest energy levels. With higher-order terms omitted, Eq. \eqref{eq:time_shift} shows that the variation of $c_k(t)$ comes from two paths: $1$-photon absorption via the interaction $U^+$ and $1$-photon emission via the interaction $U^-$. This observation allows us to make the decomposition $c_k(t)=\sum_{n=\lambda_{\text{low}}}^{\lambda_{\text{high}}} c^{(n)}_k(t)$, where $n$ is the photon absorption number, and $\lambda_{\text{high(low)}}$ the high (low) cutoff of the photon absorption number. To study the time evolution of these coefficients, we define the column vector, $C(t) = \begin{pmatrix} c^{(\lambda_{\text{high}})}(t) & \cdots & c^{(\lambda_{\text{low}})}(t) \end{pmatrix}^T$. In the time slice $[t, t+\delta{t})$, it is propagated by the tridiagonal transfer matrix
		\begin {equation}
			\begin {aligned}
			& T (t, \delta t)  \\
			=\ & \begin {pmatrix}
				\mathbbm{1} & \tilde{U}^+(t,\delta{t})\\
					\tilde{U}^-(t,\delta{t}) & \mathbbm{1} & \tilde{U}^{^+}(t,\delta{t}) \\
					& \tilde{U}^-(t,\delta{t}) & \ddots & \ddots \\
					& & \ddots & \ddots & \tilde{U}^+(t,\delta{t}) \\
					& & & \tilde{U}^-(t,\delta{t}) & \mathbbm{1} \\
				\end {pmatrix},
				\end {aligned}
			\end {equation}
		where
		\begin {equation}
			\begin {aligned}
				& \tilde{U}_{kj}^+(t,\delta{t}) = \frac{\delta{t}}{2} e^{-{\frac{t^2}{\sigma^2}}} e^{-i(E_j+\omega-E_k){t}}U^+_{kj},\\
				& \tilde{U}_{kj}^-(t,\delta{t}) = - \frac{\delta{t}}{2} e^{-{\frac{t^2}{\sigma^2}}} e^{-i(E_j-\omega-E_k){t}}U^-_{kj}.
			\end {aligned}
			\label {eq:U_tilde}
		\end {equation}
		Thus, for any arbitrary time later than $t$ by $\Delta{t}=\mathfrak{N}\delta{t}$, the quantum state of the nuclear system can be obtained by performing the successive matrix multiplications
		\begin {equation}
			C(t+M\delta{t}) = \prod_{j=\mathfrak{N}-1}^0 T(t+j\delta{t}, \delta{t})C(t).
			\label{eq:transfer_matrix_product}
		\end {equation}
        
		The probability of the $n$-photon excitation is $P^{(n)}(t)=|c^{(n)}(t)|^2$. For a system initially staying in the state $|\phi_{\text{i}}\rangle$, we fix the condition
		\begin {equation}
			c^{(0)}_k(-\infty) = \left\{ 
			\begin {aligned}
				&1\ \text{if}\ k=\text{i} \\
				&0\ \text{otherwise}
			\end {aligned}
			\right. .
		\end {equation}
		
		A special example is a monochromatic perturbation turned on at $t = 0$, this corresponds to the limit $\sigma\rightarrow+\infty$, or equivalently $e^{-t^2/\sigma^2}\rightarrow1$. In this case, we choose the initial time $t=0$. With the photon emissions are neglected by setting $U^-=0$, in the continuous limit $\delta{t}\rightarrow0$, Eq. \eqref{eq:transfer_matrix_product} recovers to a continuous form, where the $1$-photon absorption component reads
		\begin {equation}
			\begin {aligned}
				c^{(1)}_k(t) =\ & \frac{1}{2} \int_0^t d\tilde{t} e^{-i(E_{\text{i}}+\omega-E_k)\tilde{t}} U_{k\text{i}}^+ \\
				=\ & \frac{iU_{k\text{i}}}{2} \frac{e^{-i(E_{\text{i}}+\omega-E_k)t}-1}{E_{\text{i}}+\omega-E_k}
			\end {aligned}
			\label {eq:1_photon_only_absorption}
		\end {equation}
		and the $2$-photon absorption component reads
		\begin {equation}
			\begin {aligned}
				c^{(2)}_k(t) =\ & \frac{1}{2} \int_0^t d\tilde{t} e^{-i(E_j+\omega-E_k)\tilde{t}}U_{kj}^+ c^{(1)}_j(\tilde{t}) \\
				=\ & - \frac{U^+_{kj}U^+_{j\text{i}}}{4(E_i+\omega-E_j)} \\
				& \qquad \times \left[ \frac{e^{-i(E_{\text{i}}+2\omega-E_k)t}-1}{E_{\text{i}}+2\omega-E_k} \right. \\
                & \qquad \qquad \left. - \frac{e^{-i(E_j+\omega-E_k)t}-1}{E_j+\omega-E_k} \right] . \\
			\end {aligned}
			\label {eq:2_photon_only_absorption}
		\end {equation}
		This result recovers the result of G\"oppert-Mayer in Ref.~\cite{GoeppertMayer1931}.
		
	\subsection {The two-step approximation}
	
		The approximation in Eq. \eqref {eq:time_shift} requires $\Delta{E}\delta{t}\ll1$. In the case of the isomeric excitation of $^{229\text{m}}$Th, the excitation energy corresponds to a time period about $0.5\ \text{fs}$. The time slice $\delta{t}$ is required to be much smaller than this value. For long pulses around several tens of nanoseconds, with $\delta{t}=1\ \text{as}$, this requirement leads to $10^{10}$ times matrix multiplication, which makes the numerical work difficult.
		
		To reduce the computational workload, we introduce the following approximation. We firstly divide the Gaussian envelop into $\mathfrak{N}$ intervals, in the $I$-th interval we fix the time-dependent envelop $e^{-t^2/\sigma^2}$ to $e^{-t_I^2/\sigma^2}$, with $t_I$ the central value in the interval. Then in the second step, the interval is cut into $\mathfrak{n}$ slices, for the $i$-th slice in the $I$-th  interval, the transfer matrix reads
		\begin {equation}
			T _{I,i} = 
			\begin {pmatrix}
				\mathbbm{1} & \tilde{U}^+_{I,i}\\
				\tilde{U}^-_{I,i} & \mathbbm{1} & \tilde{U}^{^+}_{I,i} \\
				& \tilde{U}^-_{I,i} & \ddots & \ddots \\
				& & \ddots & \ddots & \tilde{U}^+_{I,i} \\
				& & & \tilde{U}^-_{I,i} & \mathbbm{1} \\
			\end {pmatrix},
		\end {equation}
		where
		\begin {equation}
			\begin {aligned}
				& \left[\tilde{U}_{I,i}^+\right]_{kj} = \frac{\delta{t}}{2} e^{-{\frac{t_I^2}{\sigma^2}}} e^{-i(E_j+\omega-E_k){t_i}}U^+_{kj},\\
				& \left[\tilde{U}_{I,i}^-\right]_{kj} = - \frac{\delta{t}}{2} e^{-{\frac{t_I^2}{\sigma^2}}} e^{-i(E_j-\omega-E_k){t_i}}U^-_{kj}.
			\end {aligned}
			\label {eq:U_tilde_element}
		\end {equation}
		This approximation allows us to obtain the transfer matrix of the $(i+1)$-th time slice by the transformation $T_{I, i+1} = ST_{I, i}S^{-1}$, where the matrix
        \begin {equation}
            \begin {aligned}
            S
            = \begin {pmatrix}
                \tilde{S}e^{-i\lambda_{\text{high}}\omega} & 0 \\
                0 & \tilde{S}e^{-i(\lambda_{\text{high}}-1)\omega} & 0 \\
                 & 0 & \ddots & \ddots \\
                 & & \ddots & \ddots & 0 \\
                 & & & 0 & \tilde{S}e^{-i\lambda_{\text{low}}\omega}
            \end {pmatrix},
            \end {aligned}
        \end {equation}
        contains $(\lambda_{\text{high}}-\lambda_{\text{low}}+1)\times(\lambda_{\text{high}}-\lambda_{\text{low}}+1)$ blocks. The nonzero block $\tilde{S}$ is also diagonal with $\tilde{S}_{jj}=e^{iE_j\delta{t}}$. The transfer matrix of the $I$-th interval is approximated to
		\begin {equation}
			\begin {aligned}
				T_I =\ & T_{I, n} \cdots T_{I, 2}T_{I, 1} \\
                =\ & S^{\mathfrak{n}} T_{I, 1} S^{-\mathfrak{n}} \cdots S^1T_{I, 1}S^{-1} T_{I, 1} \\
				=\ & S^{\mathfrak{n}+1} \left[S^{-1} T_{I, 1}\right]^{\mathfrak{n}}.
			\end {aligned}
		\end {equation}
		In this way, the time of matrix multiplications are reduced from $\mathfrak{n}$ to $\log_2(\mathfrak{n})$. Hence, this approximation enables the method to deal with long pulses. This approximation vanishes when $\mathfrak{n}=1$.

\section {Numerical results and discussions}
	\label {sec:numerical_results}
	
	\subsection {Choice of physical parameters}
   
        In the present work, we analyze numerically an isolated $2$-level system which involves only the ground state of the $^{229}$Th nucleus and the isomeric state isomeric state $^{229\text{m}}$Th. Numerous efforts have been made in both theory and experiments for the relevant properties of the isomeric state $^{229\text{m}}$Th \cite{RevModPhys.49.833,Dykhne1998,PhysRevLett.122.162502,PhysRevC.73.044326,Canty1977,PhysRevC.71.044303,PhysRevLett.98.142501,Seiferle2019,Kraemer2023,PhysRevLett.132.182501, PhysRevLett.133.013201,PhysRevLett.106.223001,th229ground,PhysRevA.88.060501,Thielking2018}, such as the excitation energy, the lifetime, and the electromagnetic moments. In the present work, we adopt, for the isomeric state $^{229\text{m}}$Th, the excitation energy $E_\text{I}=8.35574\ \text{eV}$, the lifetime $\tau_{\text{I}}=1740\ \text{s}$ and the reduced transition probability $B(\text{M}1,\text{I}\rightarrow\text{G})=0.022\ \text{W.u.}$ (Weisskopf unit), which are taken from Ref.~\cite{PhysRevLett.132.182501}. For the rotations within the degenerate subspaces, we adopt $\mu_{\text{G}}=0.360\mu_{\text{N}}$ for the magnetic dipole moment of the ground state of $^{229}$Th from Refs.~\cite{PhysRevLett.106.223001,th229ground,PhysRevA.88.060501}, with $\mu_{\text{N}}$ the nuclear magneton, and the dipole moment of $^{229\text{m}}$Th $\mu_{\text{I}}=-0.37\mu_{\text{N}}$, which is derived from the ratio $\mu_{\text{I}}/\mu_{\text{G}}=-1.04$ \cite{Thielking2018}. According to our calculations, the contribution from the E2 transition is much smaller than the one from the M1 transition for the case analyzed in the present work, thus we neglect the E2 transition in the present work.

		The laser intensity of $1.1\times10^{23}\ \text{W}/\text{cm}^2$ has been achieved by the team of CoReLS \cite{Yoon:21}. The same team obtained the laser intensity $5.5\times10^{22}$ in 2019 \cite{Yoon:19}. The facility delivers laser pulses with the pulse duration of $19.6\ \text{fs}$ and the peak power of $4\ \text{PW}$. There are also a number of petawatt and exawatt lasers available or in the plan of many countries \cite{Danson2015, Danson2019,Danson_2004,Waxer2005,Tiwari2019,RevModPhys.94.045001} which may provide short laser pulses with similar or higher intensities. Furthermore, another kind of laser facilities, such as LMJ and NIF \cite{MIQUEL2020,CEA/DAM2020,Hogan_2001,Haynam2007,RevModPhys.94.045001,Cerjan2018}, can provide laser beams with a long pulse duration about $10\ \text{ns}$. The numerical calculations in the present work contain two aspects: one within the parameter space around the limit of current capabilities of laser facilities and the other around near the quantum mechanical limit of the multi-photon absorption.

        Furthermore, in the numerical calculations, we set the lower cutoff $\lambda_{\text{low}}=-10$, the higher cutoff $\lambda_{\text{high}} = +12, +13\ \text{and} +14$ for the $2$-, $3$- and $4$-photon excitations, respectively. For a laser pulse with the pulse duration (FWHM) $\Gamma_{\text{I}}$, the calculation is carried out in $t\in[-5\Gamma_{\text{I}}/\sqrt{2},+5\Gamma_{\text{I}}/\sqrt{2}]$. As the numerical results for the concerned excitation processes approach constants before $t=5\Gamma_{\text{I}}/\sqrt{2}$, we define the final excitation probability $P^{(n)}_{\text{f}}=P^{(n)}(5\Gamma_{\text{I}}/\sqrt{2})$. By varying the parameters, we ensure the convergency of the numerical results.

        In the present work, only Gaussian-shaped laser pulses have been analyzed. Since the Gaussian function has a good concentration in both the time and the frequency domains, it provides a good condition to demonstrate the features of the multi-photon excitation with the photon energy centered around $E_{\text{I}}/n$. It is worth pointing out that some laser pulse shapes may cause confusing results. For example, with the envelop $f(t) = \sin^2(at)$ defined in $t\in[-\frac{\pi}{a},+\frac{\pi}{a}]$, the $1$-photon absorption dominates even if $\omega=E_{\text{I}}/2$. A Fourier transformation shows that this shape leads to large enough components around $\omega=E_\text{I}$.

	\subsection {A typical example of the excitation}
		\label {subsec:typical_excitation}
		
        \begin {figure} [h]
			\includegraphics {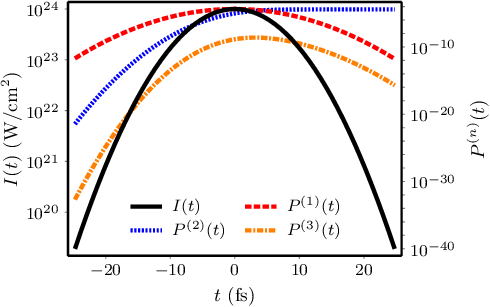}
			\caption {A typical example of the excitation with the condition $\Gamma_{\text{I}}=10\ \text{fs}$ and $I = 10^{24}\ \text{W}/\text{cm}^2$. Parameters $\mathfrak{N}=10^3$ and $\mathfrak{n}=1024$ are adopted.}
			\label {fig:typical_process}
		\end {figure}
        
        The result of a typical example of the excitation is presented in Fig. \ref{fig:typical_process}, where the laser parameters are assumed to be $\Gamma_{\text{I}}=10\ \text{fs}$ and $\omega = E_{\text{I}}/2$. Throughout the whole interaction, the $2$-photon absorption probability keeps increasing. In contrast, the $1$- and $3$- photon excitations make nonzero but negligible contribution in the final state. The cause of these nonzero probabilities is the finite width of the Gaussian pulse, which leads to small frequency components at $E_\text{I}$ and $E_\text{I}/3$.		

	\subsection {The excitation probability}
	
		Now we focus on the numerical results of the final excitation probability $P^{(n)}_{\text{f}}$. We assume here the photon energy $\omega=E_{\text{I}}/n$. We set $\mathfrak{N}=1000$ and $\mathfrak{n}$ flexible to ensure $\delta{t}<0.1\ \text{as}$.

    	\subsubsection {The $2$-photon excitation}
		
\begin {figure} [h]
                \includegraphics {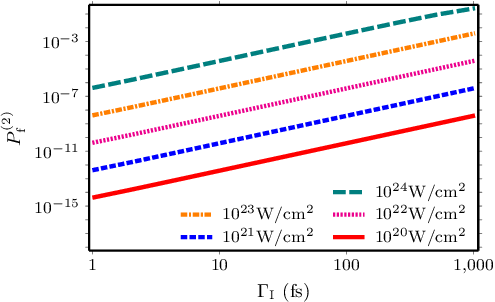}
                \caption {The $2$-photon excitation with selected peak laser intensities under the condition $I\geqslant10^{20}\ \text{W}/\text{cm}^2$.}
                \label {fig:2_photon_strong}
		      \end {figure}

\begin {figure} [h]
                \includegraphics {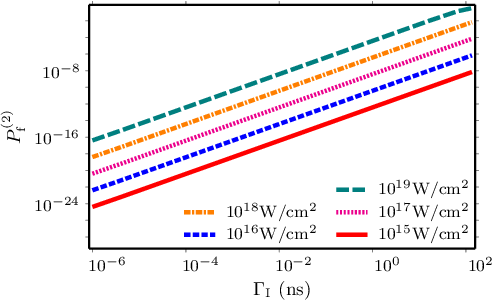}
                \caption {The $2$-photon excitation with selected peak laser intensities under the condition $I\leqslant10^{19}\ \text{W}/\text{cm}^2$.}
                \label {fig:2_photon_weak}
            \end {figure}

            In Fig. \ref{fig:2_photon_strong}, we demonstrate the relation between the $2$-photon excitation probability and the laser pulse duration (FWHM) with the intensity ranging from $10^{20}\ \text{W}/\text{cm}^2$ to $10^{24}\ \text{W}/\text{cm}^2$, assuming the photon energy $\omega=E_{\text{I}}/2$. The nuclear excitation probability is proportional to the square of the peak laser intensity. When the laser pulse duration is longer than $\sim 10$ fs, the nuclear excitation probability is also proportional to the square of the FWHM $\Gamma_{\text{I}}$. We note that, as discussed below, these relations are valid for short laser pulse durations where the leading-order process dominates the multi-photon excitation. As shown in Fig. \ref{fig:2_photon_strong}, with the laser intensity $10^{24}\ \text{W}/\text{cm}^2$, the $2$-photon excitation probability is about $10^{-1}$ for laser pulse duration FWHM of $1000\ \text{fs}$ and about $10^{-3}$ for laser pulse duration FWHM of $100\ \text{fs}$. Furthermore, considering the fact that facilities with lower laser intensity could provide laser pulses with longer pulse durations, we present in Fig. \ref{fig:2_photon_weak} the result of the $2$-photon excitation probability with the peak laser intensity ranging from $10^{15}\ \text{W}/\text{cm}^2$ to $10^{19}\ \text{W}/\text{cm}^2$, and the laser pulse duration FWHM ranging from $1\ \text{fs}$ to $100\ \text{ns}$. Again, here we assume the photon energy $\omega=E_{\text{I}}/2$.
            
            \begin {figure} [h]
                \includegraphics {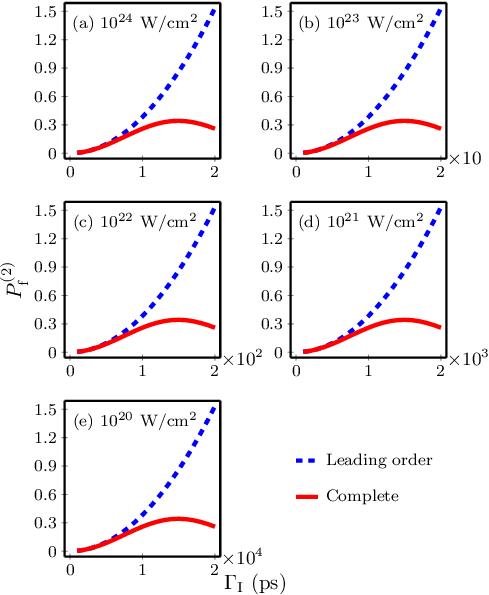}
                \caption {Leading-order and complete results of the 2-photon excitation.}
                \label {fig:2_photon_w_o_emission}
            \end {figure}
			
            For very large $t$, the monochromatic results Eq. \eqref{eq:1_photon_only_absorption} and Eq. \eqref{eq:2_photon_only_absorption} lead to unphysical probabilities larger than $1$. This is because the leading-order derivations do not include the laser-induced photon emissions. In the case of pulsed lasers, similar problems exist if the laser pulse duration is very long. By setting $U^-=0$ instead of the Hermitian conjugate of $U^+$, we can extract the leading-order contribution. The differences between the leading-order results and the complete results [where $U^{-}=\left(U^{+}\right)^{\dagger}$] are presented in Fig. \ref{fig:2_photon_w_o_emission}.  We analyze a few peak laser intensities from $I = 10^{24}\ \text{W}/\text{cm}^2$ to $I = 10^{20}\ \text{W}/\text{cm}^2$. These different intensities lead to a similar shape of the line curves, except that the less intense laser beam requires longer pulse to achieve the same probability. We can observe from Fig. \ref{fig:2_photon_w_o_emission} that the results of the $2$-photon excitation with the laser intensities considered here share a similar highest excitation probability about $0.3$. The corresponding laser pulse duration is proportional to the inverse of the peak laser intensity. It should be noted that the laser-induced photon emission become significant only when the laser pulse duration is long enough.
        
        \subsubsection {The $3$-photon excitation}
		
\begin {figure} [h]
                \includegraphics {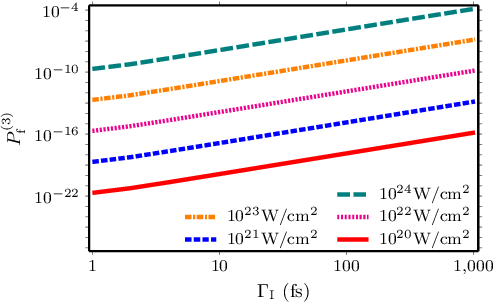}
                \caption {The $3$-photon excitation with selected peak laser intensities under the condition $I\geqslant10^{20}\ \text{W}/\text{cm}^2$.}
                \label {fig:3_photon}
            \end {figure}

            The $3$-photon excitation process is evaluated with the photon energy $\omega = E_{\text{I}} / 3$. In Fig. \ref{fig:3_photon}, we present the relation between the $3$-photon excitation probability and the laser pulse duration FWHM with selected peak laser intensities, for short laser pulse durations. Since the $3$-photon excitation process involves the 3rd or higher orders, the nuclear excitation probabilities are much smaller than the ones of the $2$-photon excitation. With the laser intensity $10^{24}\ \text{W}/\text{cm}^2$, the $3$-photon excitation probability is about $10^{-4}$ for the laser pulse duration FWHM of $1000\ \text{fs}$. It can also be observed in Fig. \ref{fig:3_photon} that, while the $3$-photon excitation probability is still proportional to the square of the laser pulse FWHM (when the laser pulse duration FWHM is longer than $\sim 10$ fs), it is proportional to the cube of the peak laser intensity, under the condition of short laser pulse durations.
            
            \begin {figure} [h]
                \includegraphics {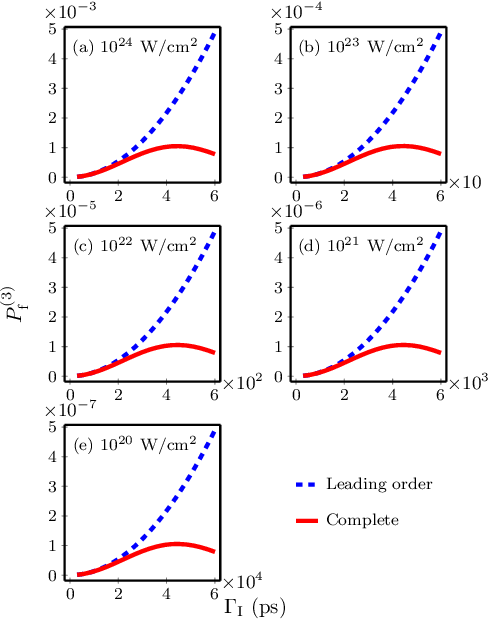}
                \caption {Leading-order and complete results of the 3-photon excitation.}
                \label {fig:3_photon_w_o_emission}
            \end {figure}

            A detailed analysis around the laser pulse duration where the higher-order effects become significant is presented in Fig. \ref{fig:3_photon_w_o_emission}. Unlike the $2$-photon excitation case, in the $3$-photon excitation, the maximum value of the nuclear excitation probability has a strong dependence on the laser intensity. This is a result from the balance between the long time evaluation of the photon absorptions and the laser-induced photon emissions.
		
		\subsubsection {The $4$-photon excitation}
     
\begin {figure} [h]
				\includegraphics {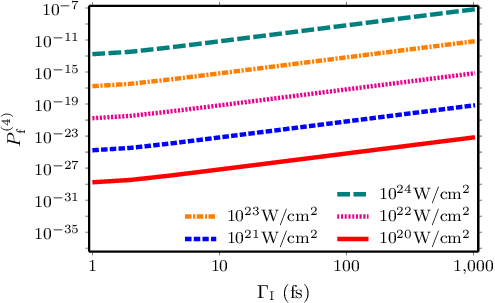}
				\caption {The $4$-photon excitation with selected peak laser intensities under the condition $I\geqslant10^{20}\ \text{W}/\text{cm}^2$.}
				\label {fig:4_photon}
			\end {figure}

            Now we turn to the case of the $4$-photon excitation, with the assumption of the photon energy $\omega = E_{\text{I}} / 4$. Figure \ref{fig:4_photon} shows the $4$-photon excitation probabilities with $I\leqslant10^{24}\ \text{W}/\text{cm}^2$ and $\Gamma_\text{I}\leqslant1000\ \text{fs}$, where the leading order process dominates. In this region, the $4$-photon excitation probability is proportional to the fourth power of the peak laser intensity and the square of the laser pulse duration FWHM (when the laser pulse duration FWHM is longer than $\sim 10$ fs). Since the $4$-photon excitation process is of the 4th or the higher orders, the corresponding excitation probabilities are much smaller than the ones of the $2$-photon excitation or the $3$-photon excitation.

			\begin {figure} [h]
				\includegraphics {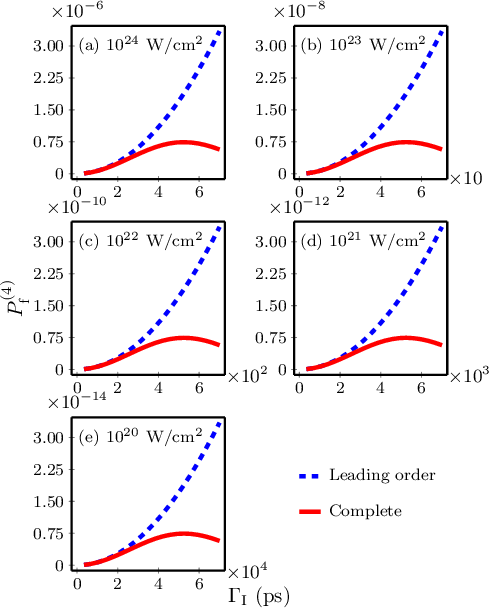}
				\caption {Leading-order and complete results of the 4-photon excitation.}
				\label {fig:4_photon_w_o_emission}
			\end {figure}

            A detailed scan around the region where the high-order effects emerge is presented in Fig. \ref{fig:4_photon_w_o_emission}. The corresponding scale of the laser pulse duration is proportional to the inverse of the peak laser intensity. Similar to the $3$-photon excitation case, the maximum probability depends on the peak laser intensity, but it is proportional to the inverse of the square of the peak laser intensity.

\section {Conclusions}
    \label {sec:conclusion}

	In this article, we have studied theoretically the multi-photon excitation to the $8$-eV isomeric state $^{229\text{m}}$Th from the ground state in the direct laser-nucleus interaction. The study has been done by solving the TDSE with a method which allows us to study the $n$-photon absorption in the nuclear excitation in the direct laser-nucleus interaction. The method enables the systematic analysis of the multi-photon absorption. Based on the laser facilities available currently or in the near future, we analyze in details the impact of the laser parameters, such as the laser intensity and the laser pulse duration, on the nuclear excitation probability of the multi-photon excitation.  Our results show that, the nuclear $n$-photon excitation probability $P^{(n)}_\text{f}$ is proportional to the square of the laser pulse duration FWHM $\Gamma_{\text{I}}^2$ and the $n$-th power of the peak laser intensity $I^n$, under the condition of short laser pulse durations where leading-order processes dominate the multi-photon excitation. Our analysis for long laser pulse durations provides the onset of the high-order effects that both the laser-induced photon absorptions and emissions play important roles for the nuclear multi-photon excitation in the direct laser-nucleus interaction. Furthermore, our results also show that high order effects set limits for the nuclear excitation probabilities. In the $2$-photon excitation, the limit of the nuclear excitation probabilities keeps to be around $0.3$ with various values of the peak laser intensity. However, in the $3$- and $4$-photon excitations, the limits increase with increasing the peak laser intensity. 

    Our results indicate that for experimental studies of the multi-photon excitation in the direct laser-nucleus interaction, high laser intensities which are similar to or even higher than the record of the achieved laser intensity in experiments are recommended. We note that although such super high laser intensities could be available currently or in the near future, such laser systems are in general have significant uncertainties on the laser parameters. Thus, currently, the case of such super high laser intensities should be impractical for the application of a nuclear clock. For the nuclear clock application, lasers with low intensities and long laser pulse durations are recommended. However, we have to emphasis that, there is a long way to go for the nuclear clock application of the multi-photon excitation in the direct laser-nucleus interaction, as there are a number of practical issues needed to be solved including a precise control of the process and the requirement of stable light sources with extremely narrow linewidths.

\begin{acknowledgements}
    This work is supported the National Natural Science Foundation of China (Grant No. 12475122), and by the Fundamental Research Funds for the Central Universities (Grant No. 010-63243088).
\end{acknowledgements}

\appendix

\section {The laser-nucleus interaction matrix}
\label {sec:laser-nucleus interaction}
	
	The direct laser-nucleus interaction is essentially the interaction between the electromagnetic field and the nucleus, so it follows the same principle as the Coulomb excitations \cite{RevModPhys.28.432, PhysRev.96.237, Biedenharn:1955zza, PhysRev.101.662}. The laser-nucleus interaction reads
	\begin {equation}
		V(t) = - \int d^3x \bs{j}\cdot\bs{A}(\bs{r},t),
		\label {eq:interaction}
	\end {equation}
	where the pulsed laser field
	\begin {equation}
		\begin {aligned}
			\bs{A}(\bs{r},t) =\ & \frac{i}{2} \bs{A}_0 e^{-t^2/\sigma^2} \left(e^{-i\omega{t}}e^{i\bs{k}\cdot\bs{r}}-e^{i\omega{t}}e^{-i\bs{k}\cdot\bs{r}}\right),
		\end {aligned}
	\end {equation}
	with $A_0=I/(\sqrt{2}k)$. This interaction leads to Eq. \eqref {eq:Schroedinger} with 
    \begin {subequations}
        \begin {align}
            & U^+ = - \int d^3x \bs{j}\cdot\bs{A}_0e^{i\bs{k}\cdot\bs{r}}, \\
            & U^- = - \int d^3x \bs{j}\cdot\bs{A}_0e^{-i\bs{k}\cdot\bs{r}}.
        \end {align}
    \end {subequations}
        
    To express the interaction Eq. \eqref{eq:interaction} in terms of multipole operators, we use the expansion \cite{Eisenberg1976,PhysRevLett.133.152503}
	\begin {equation}
		\begin {aligned}
			\hat{e}_{\nu} e^{i\bs{k}\cdot\bs{r}} =\ & \nu \sqrt{2\pi} \sum_{LM} \sqrt{2L+1} i^L D_{M\nu}^L(\phi,\theta,0) \\
			& \quad \times [A_{LM}(\bs{k};\bs{r},M) + i{\nu}A_{LM}(\bs{k};\bs{r},E)],
		\end {aligned}
		\label {eq:expansion}
	\end {equation}
	with $D_{M\nu}^L(\phi,\theta,0)$ the Wigner-D function, 
    \begin {subequations}
        \begin {align}
            & \hat{e}_{\pm1} = \mp \frac{1}{\sqrt{2}} (\hat{e}_x\pm{i}\hat{e}_y), \\
            & \hat{e}_0 = \hat{e}_z
        \end {align}
    \end {subequations}
    the spherical basis set and
	\begin {subequations}
		\begin {align}
			& A_{LM}(\bs{k};\bs{r},\text{M}) = \frac{1}{\sqrt{L(L+1)}} \bs{L} [j_L(kr)Y_{LM}(\hat{\bs{r}})], \\
			& \begin {aligned} A_{LM}(\bs{k};\bs{r},\text{E}) =\ &\frac{-i}{k\sqrt{L(L+1)}} \\
            & \qquad \nabla\times \bs{L} [j_L(kr)Y_{LM}(\hat{\bs{r}})]
            \end {aligned}
		\end {align}
	\end {subequations}
	the transverse vector spherical harmonics. With $kr\ll1$, the spherical Bessel function is approximated by $j_L(kr)\approx(kr)^L/(2L+1)!!$. For the initial case where the nucleus is not polarized, we set the direction of the laser beam as the z-axis. The operator $U^+$ is expanded as 
    \begin {equation}
        \begin {aligned} 
            U^+ =\ & A_0 \sum_L \sum_{M=\pm1} \frac{(ik)^L}{(2L+1)!!} \sqrt{\frac{(2L+1)(L+1)}{L}} \\
            & \qquad \times \left[ i\mathcal{M}(\text{M}L, M) + M\mathcal{M}(\text{E}L,M) \right],
        \end {aligned}
        \label {eq:Uplus}
    \end {equation}
	where the multipole operators are defined as
	\begin {subequations}
		\begin {align}
			& \mathcal{M}(\text{E}L,M) = \int d^3x \rho r^L Y_L^M(\hat{\bs{r}}), \\
			& \mathcal{M}(\text{M}L,M) = - \frac{i}{L+1} \int d^3x \bs{j} \cdot \bs{L} \left[r^LY_L^M(\hat{\bs{r}})\right].
		\end {align}
	\end {subequations}
        
	According to the Wigner-Eckart theorem, the matrix element of a multipole operator can be decomposed to a product of a Wigner 3j-symbol and a matrix element independent on the magnetic quantum number $M_k$ \cite{Bohr1998}
	\begin {equation}
		\begin {aligned}
			& \langle \alpha_k J_k M_k|\mathcal{M}(\tau L,M)| \alpha_j J_j M_j\rangle = (-1)^{J_k-M_k} \\
            & \qquad \qquad \times \begin{pmatrix}J_k & L & J_j \\ -M_k & M & M_j \end{pmatrix}\langle{\alpha_k}\lVert\mathcal{M}(\tau L)\rVert{\alpha_j}\rangle,
		\end {aligned}
		\label {eq:Wigner_Eckart}
	\end {equation}
	with $\tau=\text{E or M}$ denoting the electric or magnetic transition and $|\alpha_k\rangle$ denoting the subspace of degenerate states. We choose the states with explicit magnetic quantum number as the basis set $\{|\alpha_j J_j M_j\rangle\}$. Since an energy level with the angular momentum $J_\alpha$ has the degeneracy $d_\alpha=2J_\alpha+1$, a $\left(\sum_\alpha{d_{\alpha}}\right)\times\left(\sum_\alpha{d_{\alpha}}\right)$ matrix is used to express the operator $U^+$, with the elements
    \begin {equation}
        \begin {aligned}
            &\left(U^+\right)_{pq} 
             = \sum_L \sum_{M=\pm1} \frac{A_0(ik)^L}{(2L+1)!!} \\
            & \quad \times \sqrt{\frac{(2L+1)(L+1)}{L}} \\
            & \quad \times (-1)^{J_p-M_{p}} \begin{pmatrix}J_p & L & J_q \\ -M_{p} & M & M_{q} \end{pmatrix} \\
            & \quad \times \left[ i \langle \alpha_p \lVert\mathcal{M}(\text{M} L)\rVert \alpha_q \rangle + M \langle \alpha_p \lVert\mathcal{M}(\text{E} L)\rVert \alpha_q \rangle \right].
        \end {aligned}
        \label {eq:Upluselement}
    \end {equation}
    By arranging the non-degenerate states in the descending order of the energy and the degenerate states in the descending order of the magnetic quantum number, the index $q$ is given by
    \begin {equation}
        q = \sum_{E_\alpha>E_q}\left(2J_{\alpha}+1\right) + \left(J_q-M_q\right) + 1.
    \end {equation}
        
    In practice, one can alter $\alpha$ from the highest to the lowest energy level and $M_q$ from $+J_\alpha$ to $-J_\alpha$ to determine the corresponding $q$. In this arrangement, matrix representations are separated into blocks. In the off-diagonal blocks where $|\alpha_p\rangle\neq|\alpha_q\rangle$, the matrix element $\langle\alpha_p\lVert\mathcal{M}({\tau}L)\rVert\alpha_q\rangle$ is related to the reduced transition probability
	\begin {equation}
        \begin {aligned}
		& B({\tau}L;\alpha_q\rightarrow\alpha_p) \\
        & \qquad = \frac{1}{2J_{\alpha_q}+1} \left \lvert \langle\alpha_p\lVert\mathcal{M}({\tau}L) \rVert\alpha_q\rangle\right\rvert^2.
        \end {aligned}
        \label {eq:bel2mel}
	\end {equation}
	Otherwise, this matrix element is related to the multipole moments \cite{ring2004nuclear}, such as the magnetic dipole moment
	\begin {equation}
		\mu_{\alpha} = \sqrt{\frac{4\pi}{3}} \langle J_{\alpha}, J_{\alpha}|\mathcal{M}(\text{M}1,0)| J_{\alpha}, J_{\alpha} \rangle
	\end {equation}
	and the electric quadrupole moment
	\begin {equation}
		Q_{\alpha} = \sqrt{\frac{16\pi}{5}} \langle J_{\alpha}, J_{\alpha}|\mathcal{M}(\text{E}2,0)| J_{\alpha}, J_{\alpha} \rangle,
	\end {equation}
    by the Wigner-Eckart theorem Eq. \eqref{eq:Wigner_Eckart}. Hence, the matrix representation of $U^+$ is related to the properties of the nuclear states. The matrix representation of $U^-$ can be obtained by $U^-=\left(U^+\right)^\dagger$.
        
    In the present work, we study the excitation from the ground state $|\text{G}\rangle$ to the isomer $|\text{I}\rangle$. And in our numerical calculations, we assume an isolated $2$-level system which involves only the ground state of the $^{229}$Th nucleus and the isomeric state isomeric state $^{229\text{m}}$Th. Since these two levels have the angular momenta $J_{\text{I}} = \frac{3}{2}$ and $J_{\text{G}} = \frac{5}{2}$, the operator $U^+$ is represented by a $10\times10$ matrix, containing a $4$- and a $6$-dimensional subspace. Numerically, we find the contribution from M1 transition is $3$-order-of-magnitude greater than the E2 transition. Thus, only M1 transitions are included.

        \bibliographystyle{apsrev-no-url-issn}
	\bibliography{all}

\begin{thebibliography}{75}
\expandafter\ifx\csname natexlab\endcsname\relax\def\natexlab#1{#1}\fi
\expandafter\ifx\csname bibnamefont\endcsname\relax
  \def\bibnamefont#1{#1}\fi
\expandafter\ifx\csname bibfnamefont\endcsname\relax
  \def\bibfnamefont#1{#1}\fi
\expandafter\ifx\csname citenamefont\endcsname\relax
  \def\citenamefont#1{#1}\fi
\expandafter\ifx\csname url\endcsname\relax
  \def\url#1{\texttt{#1}}\fi
\expandafter\ifx\csname urlprefix\endcsname\relax\def\urlprefix{URL }\fi
\providecommand{\bibinfo}[2]{#2}
\providecommand{\eprint}[2][]{\url{#2}}

\bibitem[{\citenamefont{Göppert-Mayer}(1931)}]{GoeppertMayer1931}
\bibinfo{author}{\bibfnamefont{M.}~\bibnamefont{Göppert-Mayer}},
  \bibinfo{journal}{Annalen der Physik} \textbf{\bibinfo{volume}{401}},
  \bibinfo{pages}{273} (\bibinfo{year}{1931}).

\bibitem[{\citenamefont{Dirac}(1927)}]{Dirac243}
\bibinfo{author}{\bibfnamefont{P.~A.~M.} \bibnamefont{Dirac}},
  \bibinfo{journal}{Proceedings of the Royal Society of London A: Mathematical,
  Physical and Engineering Sciences} \textbf{\bibinfo{volume}{114}},
  \bibinfo{pages}{243} (\bibinfo{year}{1927}).

\bibitem[{\citenamefont{Kaiser and Garrett}(1961)}]{PhysRevLett.7.229}
\bibinfo{author}{\bibfnamefont{W.}~\bibnamefont{Kaiser}} \bibnamefont{and}
  \bibinfo{author}{\bibfnamefont{C.~G.~B.} \bibnamefont{Garrett}},
  \bibinfo{journal}{Phys. Rev. Lett.} \textbf{\bibinfo{volume}{7}},
  \bibinfo{pages}{229} (\bibinfo{year}{1961}).

\bibitem[{\citenamefont{Abella}(1962)}]{PhysRevLett.9.453}
\bibinfo{author}{\bibfnamefont{I.~D.} \bibnamefont{Abella}},
  \bibinfo{journal}{Phys. Rev. Lett.} \textbf{\bibinfo{volume}{9}},
  \bibinfo{pages}{453} (\bibinfo{year}{1962}).

\bibitem[{\citenamefont{Bischel
  et~al.}(1976{\natexlab{a}})\citenamefont{Bischel, Kelly, and
  Rhodes}}]{PhysRevA.13.1817}
\bibinfo{author}{\bibfnamefont{W.~K.} \bibnamefont{Bischel}},
  \bibinfo{author}{\bibfnamefont{P.~J.} \bibnamefont{Kelly}}, \bibnamefont{and}
  \bibinfo{author}{\bibfnamefont{C.~K.} \bibnamefont{Rhodes}},
  \bibinfo{journal}{Phys. Rev. A} \textbf{\bibinfo{volume}{13}},
  \bibinfo{pages}{1817} (\bibinfo{year}{1976}{\natexlab{a}}).

\bibitem[{\citenamefont{Bischel
  et~al.}(1976{\natexlab{b}})\citenamefont{Bischel, Kelly, and
  Rhodes}}]{PhysRevA.13.1829}
\bibinfo{author}{\bibfnamefont{W.~K.} \bibnamefont{Bischel}},
  \bibinfo{author}{\bibfnamefont{P.~J.} \bibnamefont{Kelly}}, \bibnamefont{and}
  \bibinfo{author}{\bibfnamefont{C.~K.} \bibnamefont{Rhodes}},
  \bibinfo{journal}{Phys. Rev. A} \textbf{\bibinfo{volume}{13}},
  \bibinfo{pages}{1829} (\bibinfo{year}{1976}{\natexlab{b}}).

\bibitem[{\citenamefont{Braunstein and Ockman}(1964)}]{PhysRev.134.A499}
\bibinfo{author}{\bibfnamefont{R.}~\bibnamefont{Braunstein}} \bibnamefont{and}
  \bibinfo{author}{\bibfnamefont{N.}~\bibnamefont{Ockman}},
  \bibinfo{journal}{Phys. Rev.} \textbf{\bibinfo{volume}{134}},
  \bibinfo{pages}{A499} (\bibinfo{year}{1964}).

\bibitem[{\citenamefont{Shimizu}(1989)}]{PhysRevB.40.1403}
\bibinfo{author}{\bibfnamefont{A.}~\bibnamefont{Shimizu}},
  \bibinfo{journal}{Phys. Rev. B} \textbf{\bibinfo{volume}{40}},
  \bibinfo{pages}{1403} (\bibinfo{year}{1989}).

\bibitem[{\citenamefont{Pattanaik et~al.}(2016)\citenamefont{Pattanaik,
  Reichert, Khurgin, Hagan, and Van~Stryland}}]{7400910}
\bibinfo{author}{\bibfnamefont{H.~S.} \bibnamefont{Pattanaik}},
  \bibinfo{author}{\bibfnamefont{M.}~\bibnamefont{Reichert}},
  \bibinfo{author}{\bibfnamefont{J.~B.} \bibnamefont{Khurgin}},
  \bibinfo{author}{\bibfnamefont{D.~J.} \bibnamefont{Hagan}}, \bibnamefont{and}
  \bibinfo{author}{\bibfnamefont{E.~W.} \bibnamefont{Van~Stryland}},
  \bibinfo{journal}{IEEE Journal of Quantum Electronics}
  \textbf{\bibinfo{volume}{52}}, \bibinfo{pages}{1} (\bibinfo{year}{2016}).

\bibitem[{\citenamefont{Larson}(2011)}]{Larson2011}
\bibinfo{author}{\bibfnamefont{A.~M.} \bibnamefont{Larson}},
  \bibinfo{journal}{Nature Photonics} \textbf{\bibinfo{volume}{5}},
  \bibinfo{pages}{1} (\bibinfo{year}{2011}).

\bibitem[{\citenamefont{Horton et~al.}(2013)\citenamefont{Horton, Wang, Kobat,
  Clark, Wise, Schaffer, and Xu}}]{Horton2013}
\bibinfo{author}{\bibfnamefont{N.~G.} \bibnamefont{Horton}},
  \bibinfo{author}{\bibfnamefont{K.}~\bibnamefont{Wang}},
  \bibinfo{author}{\bibfnamefont{D.}~\bibnamefont{Kobat}},
  \bibinfo{author}{\bibfnamefont{C.~G.} \bibnamefont{Clark}},
  \bibinfo{author}{\bibfnamefont{F.~W.} \bibnamefont{Wise}},
  \bibinfo{author}{\bibfnamefont{C.~B.} \bibnamefont{Schaffer}},
  \bibnamefont{and} \bibinfo{author}{\bibfnamefont{C.}~\bibnamefont{Xu}},
  \bibinfo{journal}{Nature Photonics} \textbf{\bibinfo{volume}{7}},
  \bibinfo{pages}{205} (\bibinfo{year}{2013}).

\bibitem[{\citenamefont{Shen et~al.}(2016)\citenamefont{Shen, Shuhendler, Ye,
  Xu, and Chen}}]{C6CS00442C}
\bibinfo{author}{\bibfnamefont{Y.}~\bibnamefont{Shen}},
  \bibinfo{author}{\bibfnamefont{A.~J.} \bibnamefont{Shuhendler}},
  \bibinfo{author}{\bibfnamefont{D.}~\bibnamefont{Ye}},
  \bibinfo{author}{\bibfnamefont{J.-J.} \bibnamefont{Xu}}, \bibnamefont{and}
  \bibinfo{author}{\bibfnamefont{H.-Y.} \bibnamefont{Chen}},
  \bibinfo{journal}{Chem. Soc. Rev.} \textbf{\bibinfo{volume}{45}},
  \bibinfo{pages}{6725} (\bibinfo{year}{2016}).

\bibitem[{\citenamefont{McKenzie et~al.}(2019)\citenamefont{McKenzie, Bryant,
  and Weinstein}}]{MCKENZIE20192}
\bibinfo{author}{\bibfnamefont{L.~K.} \bibnamefont{McKenzie}},
  \bibinfo{author}{\bibfnamefont{H.~E.} \bibnamefont{Bryant}},
  \bibnamefont{and} \bibinfo{author}{\bibfnamefont{J.~A.}
  \bibnamefont{Weinstein}}, \bibinfo{journal}{Coordination Chemistry Reviews}
  \textbf{\bibinfo{volume}{379}}, \bibinfo{pages}{2} (\bibinfo{year}{2019}),
  \bibinfo{note}{novel and Smart Photosensitizers from Molecule to
  Nanoparticle}.

\bibitem[{\citenamefont{Gu et~al.}(2017)\citenamefont{Gu, Wu, Xu, Feng, Yin,
  Chong, Qu, Yong, and Liu}}]{adma.201701076}
\bibinfo{author}{\bibfnamefont{B.}~\bibnamefont{Gu}},
  \bibinfo{author}{\bibfnamefont{W.}~\bibnamefont{Wu}},
  \bibinfo{author}{\bibfnamefont{G.}~\bibnamefont{Xu}},
  \bibinfo{author}{\bibfnamefont{G.}~\bibnamefont{Feng}},
  \bibinfo{author}{\bibfnamefont{F.}~\bibnamefont{Yin}},
  \bibinfo{author}{\bibfnamefont{P.~H.~J.} \bibnamefont{Chong}},
  \bibinfo{author}{\bibfnamefont{J.}~\bibnamefont{Qu}},
  \bibinfo{author}{\bibfnamefont{K.-T.} \bibnamefont{Yong}}, \bibnamefont{and}
  \bibinfo{author}{\bibfnamefont{B.}~\bibnamefont{Liu}},
  \bibinfo{journal}{Advanced Materials} \textbf{\bibinfo{volume}{29}},
  \bibinfo{pages}{1701076} (\bibinfo{year}{2017}).

\bibitem[{\citenamefont{Cumpston et~al.}(1999)\citenamefont{Cumpston,
  Ananthavel, Barlow, Dyer, Ehrlich, Erskine, Heikal, Kuebler, Lee,
  McCord-Maughon et~al.}}]{Cumpston1999}
\bibinfo{author}{\bibfnamefont{B.~H.} \bibnamefont{Cumpston}},
  \bibinfo{author}{\bibfnamefont{S.~P.} \bibnamefont{Ananthavel}},
  \bibinfo{author}{\bibfnamefont{S.}~\bibnamefont{Barlow}},
  \bibinfo{author}{\bibfnamefont{D.~L.} \bibnamefont{Dyer}},
  \bibinfo{author}{\bibfnamefont{J.~E.} \bibnamefont{Ehrlich}},
  \bibinfo{author}{\bibfnamefont{L.~L.} \bibnamefont{Erskine}},
  \bibinfo{author}{\bibfnamefont{A.~A.} \bibnamefont{Heikal}},
  \bibinfo{author}{\bibfnamefont{S.~M.} \bibnamefont{Kuebler}},
  \bibinfo{author}{\bibfnamefont{I.-Y.~S.} \bibnamefont{Lee}},
  \bibinfo{author}{\bibfnamefont{D.}~\bibnamefont{McCord-Maughon}},
  \bibnamefont{et~al.}, \bibinfo{journal}{Nature}
  \textbf{\bibinfo{volume}{398}}, \bibinfo{pages}{51} (\bibinfo{year}{1999}).

\bibitem[{\citenamefont{Zhang et~al.}(2017)\citenamefont{Zhang, Yue, Sun, Wang,
  Hao, and An}}]{C7TC00582B}
\bibinfo{author}{\bibfnamefont{Q.}~\bibnamefont{Zhang}},
  \bibinfo{author}{\bibfnamefont{S.}~\bibnamefont{Yue}},
  \bibinfo{author}{\bibfnamefont{H.}~\bibnamefont{Sun}},
  \bibinfo{author}{\bibfnamefont{X.}~\bibnamefont{Wang}},
  \bibinfo{author}{\bibfnamefont{X.}~\bibnamefont{Hao}}, \bibnamefont{and}
  \bibinfo{author}{\bibfnamefont{S.}~\bibnamefont{An}}, \bibinfo{journal}{J.
  Mater. Chem. C} \textbf{\bibinfo{volume}{5}}, \bibinfo{pages}{3838}
  (\bibinfo{year}{2017}).

\bibitem[{\citenamefont{Thielking et~al.}(2023)\citenamefont{Thielking, Zhang,
  Tiedau, Zander, Zitzer, Okhapkin, and Peik}}]{Thielking2023}
\bibinfo{author}{\bibfnamefont{J.}~\bibnamefont{Thielking}},
  \bibinfo{author}{\bibfnamefont{K.}~\bibnamefont{Zhang}},
  \bibinfo{author}{\bibfnamefont{J.}~\bibnamefont{Tiedau}},
  \bibinfo{author}{\bibfnamefont{J.}~\bibnamefont{Zander}},
  \bibinfo{author}{\bibfnamefont{G.}~\bibnamefont{Zitzer}},
  \bibinfo{author}{\bibfnamefont{M.~V.} \bibnamefont{Okhapkin}},
  \bibnamefont{and} \bibinfo{author}{\bibfnamefont{E.}~\bibnamefont{Peik}},
  \bibinfo{journal}{New Journal of Physics} \textbf{\bibinfo{volume}{25}},
  \bibinfo{pages}{083026} (\bibinfo{year}{2023}).

\bibitem[{\citenamefont{Berdah et~al.}(1996)\citenamefont{Berdah, Visticot,
  Dedonder-Lardeux, Solgadi, and Soep}}]{BERDAH1996118}
\bibinfo{author}{\bibfnamefont{M.}~\bibnamefont{Berdah}},
  \bibinfo{author}{\bibfnamefont{J.}~\bibnamefont{Visticot}},
  \bibinfo{author}{\bibfnamefont{C.}~\bibnamefont{Dedonder-Lardeux}},
  \bibinfo{author}{\bibfnamefont{D.}~\bibnamefont{Solgadi}}, \bibnamefont{and}
  \bibinfo{author}{\bibfnamefont{B.}~\bibnamefont{Soep}},
  \bibinfo{journal}{Optics Communications} \textbf{\bibinfo{volume}{124}},
  \bibinfo{pages}{118} (\bibinfo{year}{1996}).

\bibitem[{\citenamefont{Bjorklund}(1975)}]{1068619}
\bibinfo{author}{\bibfnamefont{G.}~\bibnamefont{Bjorklund}},
  \bibinfo{journal}{IEEE Journal of Quantum Electronics}
  \textbf{\bibinfo{volume}{11}}, \bibinfo{pages}{287} (\bibinfo{year}{1975}).

\bibitem[{\citenamefont{Collins
  et~al.}(1979{\natexlab{a}})\citenamefont{Collins, Olariu, Petrascu, and
  Popescu}}]{PhysRevLett.42.1397}
\bibinfo{author}{\bibfnamefont{C.~B.} \bibnamefont{Collins}},
  \bibinfo{author}{\bibfnamefont{S.}~\bibnamefont{Olariu}},
  \bibinfo{author}{\bibfnamefont{M.}~\bibnamefont{Petrascu}}, \bibnamefont{and}
  \bibinfo{author}{\bibfnamefont{I.}~\bibnamefont{Popescu}},
  \bibinfo{journal}{Phys. Rev. Lett.} \textbf{\bibinfo{volume}{42}},
  \bibinfo{pages}{1397} (\bibinfo{year}{1979}{\natexlab{a}}).

\bibitem[{\citenamefont{Collins
  et~al.}(1979{\natexlab{b}})\citenamefont{Collins, Olariu, Petrascu, and
  Popescu}}]{PhysRevC.20.1942}
\bibinfo{author}{\bibfnamefont{C.~B.} \bibnamefont{Collins}},
  \bibinfo{author}{\bibfnamefont{S.}~\bibnamefont{Olariu}},
  \bibinfo{author}{\bibfnamefont{M.}~\bibnamefont{Petrascu}}, \bibnamefont{and}
  \bibinfo{author}{\bibfnamefont{I.}~\bibnamefont{Popescu}},
  \bibinfo{journal}{Phys. Rev. C} \textbf{\bibinfo{volume}{20}},
  \bibinfo{pages}{1942} (\bibinfo{year}{1979}{\natexlab{b}}).

\bibitem[{\citenamefont{Olariu et~al.}(1981)\citenamefont{Olariu, Popescu, and
  Collins}}]{PhysRevC.23.50}
\bibinfo{author}{\bibfnamefont{S.}~\bibnamefont{Olariu}},
  \bibinfo{author}{\bibfnamefont{I.}~\bibnamefont{Popescu}}, \bibnamefont{and}
  \bibinfo{author}{\bibfnamefont{C.~B.} \bibnamefont{Collins}},
  \bibinfo{journal}{Phys. Rev. C} \textbf{\bibinfo{volume}{23}},
  \bibinfo{pages}{50} (\bibinfo{year}{1981}).

\bibitem[{\citenamefont{Yang et~al.}(2024)\citenamefont{Yang, Spohr, Cernaianu,
  Doria, Ghenuche, and Horny}}]{yang2024new}
\bibinfo{author}{\bibfnamefont{C.~J.} \bibnamefont{Yang}},
  \bibinfo{author}{\bibfnamefont{K.~M.} \bibnamefont{Spohr}},
  \bibinfo{author}{\bibfnamefont{M.}~\bibnamefont{Cernaianu}},
  \bibinfo{author}{\bibfnamefont{D.}~\bibnamefont{Doria}},
  \bibinfo{author}{\bibfnamefont{P.}~\bibnamefont{Ghenuche}}, \bibnamefont{and}
  \bibinfo{author}{\bibfnamefont{V.}~\bibnamefont{Horny}},
  \emph{\bibinfo{title}{A new scheme for isomer pumping and depletion with
  high-power lasers}} (\bibinfo{year}{2024}), \eprint{arXiv: 2404.07909}.

\bibitem[{\citenamefont{Bilous}(2018)}]{item_3026914}
\bibinfo{author}{\bibfnamefont{P.}~\bibnamefont{Bilous}}, Ph.D. thesis,
  \bibinfo{school}{Ruprecht-Karls Universit{\"a}t},
  \bibinfo{address}{Heidelberg} (\bibinfo{year}{2018}).

\bibitem[{\citenamefont{Zhang et~al.}(2024{\natexlab{a}})\citenamefont{Zhang,
  Li, and Wang}}]{PhysRevLett.133.152503}
\bibinfo{author}{\bibfnamefont{H.}~\bibnamefont{Zhang}},
  \bibinfo{author}{\bibfnamefont{T.}~\bibnamefont{Li}}, \bibnamefont{and}
  \bibinfo{author}{\bibfnamefont{X.}~\bibnamefont{Wang}},
  \bibinfo{journal}{Phys. Rev. Lett.} \textbf{\bibinfo{volume}{133}},
  \bibinfo{pages}{152503} (\bibinfo{year}{2024}{\natexlab{a}}).

\bibitem[{\citenamefont{Lu et~al.}(2025)\citenamefont{Lu, Zhang, Li, Ababekri,
  Wang, and Li}}]{lu2025nuclearexcitationcontrolinduced}
\bibinfo{author}{\bibfnamefont{Z.-W.} \bibnamefont{Lu}},
  \bibinfo{author}{\bibfnamefont{H.}~\bibnamefont{Zhang}},
  \bibinfo{author}{\bibfnamefont{T.}~\bibnamefont{Li}},
  \bibinfo{author}{\bibfnamefont{M.}~\bibnamefont{Ababekri}},
  \bibinfo{author}{\bibfnamefont{X.}~\bibnamefont{Wang}}, \bibnamefont{and}
  \bibinfo{author}{\bibfnamefont{J.-X.} \bibnamefont{Li}},
  \emph{\bibinfo{title}{Nuclear excitation and control induced by intense
  vortex laser}} (\bibinfo{year}{2025}), \eprint{arXiv: 2503.12812}.

\bibitem[{\citenamefont{von~der Wense et~al.}(2020)\citenamefont{von~der Wense,
  Bilous, Seiferle, Stellmer, Weitenberg, Thirolf, Pálffy, and
  Kazakov}}]{Wense2020a}
\bibinfo{author}{\bibfnamefont{L.}~\bibnamefont{von~der Wense}},
  \bibinfo{author}{\bibfnamefont{P.~V.} \bibnamefont{Bilous}},
  \bibinfo{author}{\bibfnamefont{B.}~\bibnamefont{Seiferle}},
  \bibinfo{author}{\bibfnamefont{S.}~\bibnamefont{Stellmer}},
  \bibinfo{author}{\bibfnamefont{J.}~\bibnamefont{Weitenberg}},
  \bibinfo{author}{\bibfnamefont{P.~G.} \bibnamefont{Thirolf}},
  \bibinfo{author}{\bibfnamefont{A.}~\bibnamefont{Pálffy}}, \bibnamefont{and}
  \bibinfo{author}{\bibfnamefont{G.}~\bibnamefont{Kazakov}},
  \bibinfo{journal}{The European Physical Journal A}
  \textbf{\bibinfo{volume}{56}}, \bibinfo{pages}{176} (\bibinfo{year}{2020}).

\bibitem[{\citenamefont{von~der Wense and Seiferle}(2020)}]{Wense2020}
\bibinfo{author}{\bibfnamefont{L.}~\bibnamefont{von~der Wense}}
  \bibnamefont{and} \bibinfo{author}{\bibfnamefont{B.}~\bibnamefont{Seiferle}},
  \bibinfo{journal}{The European Physical Journal A}
  \textbf{\bibinfo{volume}{56}}, \bibinfo{pages}{277} (\bibinfo{year}{2020}).

\bibitem[{\citenamefont{Kazakov et~al.}(2012)\citenamefont{Kazakov, Litvinov,
  Romanenko, Yatsenko, Romanenko, Schreitl, Winkler, and Schumm}}]{Kazakov2012}
\bibinfo{author}{\bibfnamefont{G.~A.} \bibnamefont{Kazakov}},
  \bibinfo{author}{\bibfnamefont{A.~N.} \bibnamefont{Litvinov}},
  \bibinfo{author}{\bibfnamefont{V.~I.} \bibnamefont{Romanenko}},
  \bibinfo{author}{\bibfnamefont{L.~P.} \bibnamefont{Yatsenko}},
  \bibinfo{author}{\bibfnamefont{A.~V.} \bibnamefont{Romanenko}},
  \bibinfo{author}{\bibfnamefont{M.}~\bibnamefont{Schreitl}},
  \bibinfo{author}{\bibfnamefont{G.}~\bibnamefont{Winkler}}, \bibnamefont{and}
  \bibinfo{author}{\bibfnamefont{T.}~\bibnamefont{Schumm}},
  \bibinfo{journal}{New Journal of Physics} \textbf{\bibinfo{volume}{14}},
  \bibinfo{pages}{083019} (\bibinfo{year}{2012}).

\bibitem[{\citenamefont{Thirolf et~al.}(2019)\citenamefont{Thirolf, Seiferle,
  and von~der Wense}}]{Thirolf_2019}
\bibinfo{author}{\bibfnamefont{P.~G.} \bibnamefont{Thirolf}},
  \bibinfo{author}{\bibfnamefont{B.}~\bibnamefont{Seiferle}}, \bibnamefont{and}
  \bibinfo{author}{\bibfnamefont{L.}~\bibnamefont{von~der Wense}},
  \bibinfo{journal}{Journal of Physics B: Atomic, Molecular and Optical
  Physics} \textbf{\bibinfo{volume}{52}}, \bibinfo{pages}{203001}
  (\bibinfo{year}{2019}).

\bibitem[{\citenamefont{Beeks et~al.}(2021)\citenamefont{Beeks, Sikorsky,
  Schumm, Thielking, Okhapkin, and Peik}}]{Beeks2021}
\bibinfo{author}{\bibfnamefont{K.}~\bibnamefont{Beeks}},
  \bibinfo{author}{\bibfnamefont{T.}~\bibnamefont{Sikorsky}},
  \bibinfo{author}{\bibfnamefont{T.}~\bibnamefont{Schumm}},
  \bibinfo{author}{\bibfnamefont{J.}~\bibnamefont{Thielking}},
  \bibinfo{author}{\bibfnamefont{M.~V.} \bibnamefont{Okhapkin}},
  \bibnamefont{and} \bibinfo{author}{\bibfnamefont{E.}~\bibnamefont{Peik}},
  \bibinfo{journal}{Nature Reviews Physics} \textbf{\bibinfo{volume}{3}},
  \bibinfo{pages}{238} (\bibinfo{year}{2021}).

\bibitem[{\citenamefont{Peik et~al.}(2021)\citenamefont{Peik, Schumm,
  Safronova, Pálffy, Weitenberg, and Thirolf}}]{Peik2021}
\bibinfo{author}{\bibfnamefont{E.}~\bibnamefont{Peik}},
  \bibinfo{author}{\bibfnamefont{T.}~\bibnamefont{Schumm}},
  \bibinfo{author}{\bibfnamefont{M.~S.} \bibnamefont{Safronova}},
  \bibinfo{author}{\bibfnamefont{A.}~\bibnamefont{Pálffy}},
  \bibinfo{author}{\bibfnamefont{J.}~\bibnamefont{Weitenberg}},
  \bibnamefont{and} \bibinfo{author}{\bibfnamefont{P.~G.}
  \bibnamefont{Thirolf}}, \bibinfo{journal}{Quantum Science and Technology}
  \textbf{\bibinfo{volume}{6}}, \bibinfo{pages}{034002} (\bibinfo{year}{2021}).

\bibitem[{\citenamefont{Tiedau et~al.}(2024)\citenamefont{Tiedau, Okhapkin,
  Zhang, Thielking, Zitzer, Peik, Schaden, Pronebner, Morawetz, De~Col
  et~al.}}]{PhysRevLett.132.182501}
\bibinfo{author}{\bibfnamefont{J.}~\bibnamefont{Tiedau}},
  \bibinfo{author}{\bibfnamefont{M.~V.} \bibnamefont{Okhapkin}},
  \bibinfo{author}{\bibfnamefont{K.}~\bibnamefont{Zhang}},
  \bibinfo{author}{\bibfnamefont{J.}~\bibnamefont{Thielking}},
  \bibinfo{author}{\bibfnamefont{G.}~\bibnamefont{Zitzer}},
  \bibinfo{author}{\bibfnamefont{E.}~\bibnamefont{Peik}},
  \bibinfo{author}{\bibfnamefont{F.}~\bibnamefont{Schaden}},
  \bibinfo{author}{\bibfnamefont{T.}~\bibnamefont{Pronebner}},
  \bibinfo{author}{\bibfnamefont{I.}~\bibnamefont{Morawetz}},
  \bibinfo{author}{\bibfnamefont{L.~T.} \bibnamefont{De~Col}},
  \bibnamefont{et~al.}, \bibinfo{journal}{Phys. Rev. Lett.}
  \textbf{\bibinfo{volume}{132}}, \bibinfo{pages}{182501}
  (\bibinfo{year}{2024}).

\bibitem[{\citenamefont{Elwell et~al.}(2024)\citenamefont{Elwell, Schneider,
  Jeet, Terhune, Morgan, Alexandrova, Tran~Tan, Derevianko, and
  Hudson}}]{PhysRevLett.133.013201}
\bibinfo{author}{\bibfnamefont{R.}~\bibnamefont{Elwell}},
  \bibinfo{author}{\bibfnamefont{C.}~\bibnamefont{Schneider}},
  \bibinfo{author}{\bibfnamefont{J.}~\bibnamefont{Jeet}},
  \bibinfo{author}{\bibfnamefont{J.~E.~S.} \bibnamefont{Terhune}},
  \bibinfo{author}{\bibfnamefont{H.~W.~T.} \bibnamefont{Morgan}},
  \bibinfo{author}{\bibfnamefont{A.~N.} \bibnamefont{Alexandrova}},
  \bibinfo{author}{\bibfnamefont{H.~B.} \bibnamefont{Tran~Tan}},
  \bibinfo{author}{\bibfnamefont{A.}~\bibnamefont{Derevianko}},
  \bibnamefont{and} \bibinfo{author}{\bibfnamefont{E.~R.}
  \bibnamefont{Hudson}}, \bibinfo{journal}{Phys. Rev. Lett.}
  \textbf{\bibinfo{volume}{133}}, \bibinfo{pages}{013201}
  (\bibinfo{year}{2024}).

\bibitem[{\citenamefont{Zhang et~al.}(2024{\natexlab{b}})\citenamefont{Zhang,
  Ooi, Higgins, Doyle, von~der Wense, Beeks, Leitner, Kazakov, Li, Thirolf
  et~al.}}]{Zhang2024}
\bibinfo{author}{\bibfnamefont{C.}~\bibnamefont{Zhang}},
  \bibinfo{author}{\bibfnamefont{T.}~\bibnamefont{Ooi}},
  \bibinfo{author}{\bibfnamefont{J.~S.} \bibnamefont{Higgins}},
  \bibinfo{author}{\bibfnamefont{J.~F.} \bibnamefont{Doyle}},
  \bibinfo{author}{\bibfnamefont{L.}~\bibnamefont{von~der Wense}},
  \bibinfo{author}{\bibfnamefont{K.}~\bibnamefont{Beeks}},
  \bibinfo{author}{\bibfnamefont{A.}~\bibnamefont{Leitner}},
  \bibinfo{author}{\bibfnamefont{G.~A.} \bibnamefont{Kazakov}},
  \bibinfo{author}{\bibfnamefont{P.}~\bibnamefont{Li}},
  \bibinfo{author}{\bibfnamefont{P.~G.} \bibnamefont{Thirolf}},
  \bibnamefont{et~al.}, \bibinfo{journal}{Nature}
  \textbf{\bibinfo{volume}{633}}, \bibinfo{pages}{63}
  (\bibinfo{year}{2024}{\natexlab{b}}).

\bibitem[{\citenamefont{Rumi and Perry}(2010)}]{Rumi2010}
\bibinfo{author}{\bibfnamefont{M.}~\bibnamefont{Rumi}} \bibnamefont{and}
  \bibinfo{author}{\bibfnamefont{J.~W.} \bibnamefont{Perry}},
  \bibinfo{journal}{Adv. Opt. Photon.} \textbf{\bibinfo{volume}{2}},
  \bibinfo{pages}{451} (\bibinfo{year}{2010}).

\bibitem[{\citenamefont{He et~al.}(2002)\citenamefont{He, Lin, Prasad, Kannan,
  Vaia, and Tan}}]{He2002}
\bibinfo{author}{\bibfnamefont{G.~S.} \bibnamefont{He}},
  \bibinfo{author}{\bibfnamefont{T.-C.} \bibnamefont{Lin}},
  \bibinfo{author}{\bibfnamefont{P.~N.} \bibnamefont{Prasad}},
  \bibinfo{author}{\bibfnamefont{R.}~\bibnamefont{Kannan}},
  \bibinfo{author}{\bibfnamefont{R.~A.} \bibnamefont{Vaia}}, \bibnamefont{and}
  \bibinfo{author}{\bibfnamefont{L.-S.} \bibnamefont{Tan}},
  \bibinfo{journal}{Opt. Express} \textbf{\bibinfo{volume}{10}},
  \bibinfo{pages}{566} (\bibinfo{year}{2002}).

\bibitem[{\citenamefont{Zelevinsky}(2010)}]{Zelevinsky2010}
\bibinfo{author}{\bibfnamefont{V.}~\bibnamefont{Zelevinsky}},
  \emph{\bibinfo{title}{Quantum Physics: Volume 2 - From Time-Dependent
  Dynamics to Many-Body Physics and Quantum Chaos}}, Physics textbook
  (\bibinfo{publisher}{Wiley}, \bibinfo{year}{2010}), ISBN
  \bibinfo{isbn}{9783527409846}.

\bibitem[{\citenamefont{Yoon et~al.}(2021)\citenamefont{Yoon, Kim, Choi, Sung,
  Lee, Lee, and Nam}}]{Yoon:21}
\bibinfo{author}{\bibfnamefont{J.~W.} \bibnamefont{Yoon}},
  \bibinfo{author}{\bibfnamefont{Y.~G.} \bibnamefont{Kim}},
  \bibinfo{author}{\bibfnamefont{I.~W.} \bibnamefont{Choi}},
  \bibinfo{author}{\bibfnamefont{J.~H.} \bibnamefont{Sung}},
  \bibinfo{author}{\bibfnamefont{H.~W.} \bibnamefont{Lee}},
  \bibinfo{author}{\bibfnamefont{S.~K.} \bibnamefont{Lee}}, \bibnamefont{and}
  \bibinfo{author}{\bibfnamefont{C.~H.} \bibnamefont{Nam}},
  \bibinfo{journal}{Optica} \textbf{\bibinfo{volume}{8}}, \bibinfo{pages}{630}
  (\bibinfo{year}{2021}).

\bibitem[{\citenamefont{Danson et~al.}(2015)\citenamefont{Danson, Hillier,
  Hopps, and Neely}}]{Danson2015}
\bibinfo{author}{\bibfnamefont{C.}~\bibnamefont{Danson}},
  \bibinfo{author}{\bibfnamefont{D.}~\bibnamefont{Hillier}},
  \bibinfo{author}{\bibfnamefont{N.}~\bibnamefont{Hopps}}, \bibnamefont{and}
  \bibinfo{author}{\bibfnamefont{D.}~\bibnamefont{Neely}},
  \bibinfo{journal}{High Power Laser Science and Engineering}
  \textbf{\bibinfo{volume}{3}}, \bibinfo{pages}{e3} (\bibinfo{year}{2015}).

\bibitem[{\citenamefont{Danson et~al.}(2019)\citenamefont{Danson, Haefner,
  Bromage, Butcher, Chanteloup, Chowdhury, Galvanauskas, Gizzi, Hein, Hillier
  et~al.}}]{Danson2019}
\bibinfo{author}{\bibfnamefont{C.~N.} \bibnamefont{Danson}},
  \bibinfo{author}{\bibfnamefont{C.}~\bibnamefont{Haefner}},
  \bibinfo{author}{\bibfnamefont{J.}~\bibnamefont{Bromage}},
  \bibinfo{author}{\bibfnamefont{T.}~\bibnamefont{Butcher}},
  \bibinfo{author}{\bibfnamefont{J.-C.~F.} \bibnamefont{Chanteloup}},
  \bibinfo{author}{\bibfnamefont{E.~A.} \bibnamefont{Chowdhury}},
  \bibinfo{author}{\bibfnamefont{A.}~\bibnamefont{Galvanauskas}},
  \bibinfo{author}{\bibfnamefont{L.~A.} \bibnamefont{Gizzi}},
  \bibinfo{author}{\bibfnamefont{J.}~\bibnamefont{Hein}},
  \bibinfo{author}{\bibfnamefont{D.~I.} \bibnamefont{Hillier}},
  \bibnamefont{et~al.}, \bibinfo{journal}{High Power Laser Science and
  Engineering} \textbf{\bibinfo{volume}{7}}, \bibinfo{pages}{e54}
  (\bibinfo{year}{2019}).

\bibitem[{\citenamefont{Gonoskov et~al.}(2022)\citenamefont{Gonoskov,
  Blackburn, Marklund, and Bulanov}}]{RevModPhys.94.045001}
\bibinfo{author}{\bibfnamefont{A.}~\bibnamefont{Gonoskov}},
  \bibinfo{author}{\bibfnamefont{T.~G.} \bibnamefont{Blackburn}},
  \bibinfo{author}{\bibfnamefont{M.}~\bibnamefont{Marklund}}, \bibnamefont{and}
  \bibinfo{author}{\bibfnamefont{S.~S.} \bibnamefont{Bulanov}},
  \bibinfo{journal}{Rev. Mod. Phys.} \textbf{\bibinfo{volume}{94}},
  \bibinfo{pages}{045001} (\bibinfo{year}{2022}).

\bibitem[{\citenamefont{MIQUEL et~al.}(2020)\citenamefont{MIQUEL, BATANI, and
  BLANCHOT}}]{MIQUEL2020}
\bibinfo{author}{\bibfnamefont{J.-L.} \bibnamefont{MIQUEL}},
  \bibinfo{author}{\bibfnamefont{D.}~\bibnamefont{BATANI}}, \bibnamefont{and}
  \bibinfo{author}{\bibfnamefont{N.}~\bibnamefont{BLANCHOT}},
  \bibinfo{journal}{The Review of Laser Engineering}
  \textbf{\bibinfo{volume}{42}}, \bibinfo{pages}{131} (\bibinfo{year}{2020}).

\bibitem[{CEA()}]{CEA/DAM2020}
\bibinfo{note}{See
  \url{http://www.asso-alp.fr/wp-content/uploads/2020/10/LMJ-PETAL-Users-guide-v2.0.pdf}}.

\bibitem[{\citenamefont{Hogan et~al.}(2001)\citenamefont{Hogan, Moses, Warner,
  Sorem, and Soures}}]{Hogan_2001}
\bibinfo{author}{\bibfnamefont{W.}~\bibnamefont{Hogan}},
  \bibinfo{author}{\bibfnamefont{E.}~\bibnamefont{Moses}},
  \bibinfo{author}{\bibfnamefont{B.}~\bibnamefont{Warner}},
  \bibinfo{author}{\bibfnamefont{M.}~\bibnamefont{Sorem}}, \bibnamefont{and}
  \bibinfo{author}{\bibfnamefont{J.}~\bibnamefont{Soures}},
  \bibinfo{journal}{Nuclear Fusion} \textbf{\bibinfo{volume}{41}},
  \bibinfo{pages}{567} (\bibinfo{year}{2001}).

\bibitem[{\citenamefont{Haynam et~al.}(2007)\citenamefont{Haynam, Wegner,
  Auerbach, Bowers, Dixit, Erbert, Heestand, Henesian, Hermann, Jancaitis
  et~al.}}]{Haynam2007}
\bibinfo{author}{\bibfnamefont{C.~A.} \bibnamefont{Haynam}},
  \bibinfo{author}{\bibfnamefont{P.~J.} \bibnamefont{Wegner}},
  \bibinfo{author}{\bibfnamefont{J.~M.} \bibnamefont{Auerbach}},
  \bibinfo{author}{\bibfnamefont{M.~W.} \bibnamefont{Bowers}},
  \bibinfo{author}{\bibfnamefont{S.~N.} \bibnamefont{Dixit}},
  \bibinfo{author}{\bibfnamefont{G.~V.} \bibnamefont{Erbert}},
  \bibinfo{author}{\bibfnamefont{G.~M.} \bibnamefont{Heestand}},
  \bibinfo{author}{\bibfnamefont{M.~A.} \bibnamefont{Henesian}},
  \bibinfo{author}{\bibfnamefont{M.~R.} \bibnamefont{Hermann}},
  \bibinfo{author}{\bibfnamefont{K.~S.} \bibnamefont{Jancaitis}},
  \bibnamefont{et~al.}, \bibinfo{journal}{Appl. Opt.}
  \textbf{\bibinfo{volume}{46}}, \bibinfo{pages}{3276} (\bibinfo{year}{2007}).

\bibitem[{\citenamefont{Cerjan et~al.}(2018)\citenamefont{Cerjan, Bernstein,
  Hopkins, Bionta, Bleuel, Caggiano, Cassata, Brune, Fittinghoff, Frenje
  et~al.}}]{Cerjan2018}
\bibinfo{author}{\bibfnamefont{C.~J.} \bibnamefont{Cerjan}},
  \bibinfo{author}{\bibfnamefont{L.}~\bibnamefont{Bernstein}},
  \bibinfo{author}{\bibfnamefont{L.~B.} \bibnamefont{Hopkins}},
  \bibinfo{author}{\bibfnamefont{R.~M.} \bibnamefont{Bionta}},
  \bibinfo{author}{\bibfnamefont{D.~L.} \bibnamefont{Bleuel}},
  \bibinfo{author}{\bibfnamefont{J.~A.} \bibnamefont{Caggiano}},
  \bibinfo{author}{\bibfnamefont{W.~S.} \bibnamefont{Cassata}},
  \bibinfo{author}{\bibfnamefont{C.~R.} \bibnamefont{Brune}},
  \bibinfo{author}{\bibfnamefont{D.}~\bibnamefont{Fittinghoff}},
  \bibinfo{author}{\bibfnamefont{J.}~\bibnamefont{Frenje}},
  \bibnamefont{et~al.}, \bibinfo{journal}{Journal of Physics G: Nuclear and
  Particle Physics} \textbf{\bibinfo{volume}{45}}, \bibinfo{pages}{033003}
  (\bibinfo{year}{2018}).

\bibitem[{\citenamefont{van Dijk et~al.}(2011)\citenamefont{van Dijk, Brown,
  and Spyksma}}]{PhysRevE.84.056703}
\bibinfo{author}{\bibfnamefont{W.}~\bibnamefont{van Dijk}},
  \bibinfo{author}{\bibfnamefont{J.}~\bibnamefont{Brown}}, \bibnamefont{and}
  \bibinfo{author}{\bibfnamefont{K.}~\bibnamefont{Spyksma}},
  \bibinfo{journal}{Phys. Rev. E} \textbf{\bibinfo{volume}{84}},
  \bibinfo{pages}{056703} (\bibinfo{year}{2011}).

\bibitem[{\citenamefont{Pic\'on}(2017)}]{PhysRevA.95.023401}
\bibinfo{author}{\bibfnamefont{A.}~\bibnamefont{Pic\'on}},
  \bibinfo{journal}{Phys. Rev. A} \textbf{\bibinfo{volume}{95}},
  \bibinfo{pages}{023401} (\bibinfo{year}{2017}).

\bibitem[{\citenamefont{Tong and Toshima}(2011)}]{Tong2011}
\bibinfo{author}{\bibfnamefont{X.-M.} \bibnamefont{Tong}} \bibnamefont{and}
  \bibinfo{author}{\bibfnamefont{N.}~\bibnamefont{Toshima}},
  \bibinfo{journal}{Computer Physics Communications}
  \textbf{\bibinfo{volume}{182}}, \bibinfo{pages}{21} (\bibinfo{year}{2011}),
  \bibinfo{note}{computer Physics Communications Special Edition for Conference
  on Computational Physics Kaohsiung, Taiwan, Dec 15-19, 2009}.

\bibitem[{\citenamefont{Zakavi and Sabaeian}(2024)}]{Zakavi2024}
\bibinfo{author}{\bibfnamefont{M.}~\bibnamefont{Zakavi}} \bibnamefont{and}
  \bibinfo{author}{\bibfnamefont{M.}~\bibnamefont{Sabaeian}},
  \bibinfo{journal}{Scientific Reports} \textbf{\bibinfo{volume}{14}},
  \bibinfo{pages}{9005} (\bibinfo{year}{2024}).

\bibitem[{\citenamefont{Chasman et~al.}(1977)\citenamefont{Chasman, Ahmad,
  Friedman, and Erskine}}]{RevModPhys.49.833}
\bibinfo{author}{\bibfnamefont{R.~R.} \bibnamefont{Chasman}},
  \bibinfo{author}{\bibfnamefont{I.}~\bibnamefont{Ahmad}},
  \bibinfo{author}{\bibfnamefont{A.~M.} \bibnamefont{Friedman}},
  \bibnamefont{and} \bibinfo{author}{\bibfnamefont{J.~R.}
  \bibnamefont{Erskine}}, \bibinfo{journal}{Rev. Mod. Phys.}
  \textbf{\bibinfo{volume}{49}}, \bibinfo{pages}{833} (\bibinfo{year}{1977}).

\bibitem[{\citenamefont{Dykhne and Tkalya}(1998)}]{Dykhne1998}
\bibinfo{author}{\bibfnamefont{A.~M.} \bibnamefont{Dykhne}} \bibnamefont{and}
  \bibinfo{author}{\bibfnamefont{E.~V.} \bibnamefont{Tkalya}},
  \bibinfo{journal}{Journal of Experimental and Theoretical Physics Letters}
  \textbf{\bibinfo{volume}{67}}, \bibinfo{pages}{251} (\bibinfo{year}{1998}).

\bibitem[{\citenamefont{Minkov and P\'alffy}(2019)}]{PhysRevLett.122.162502}
\bibinfo{author}{\bibfnamefont{N.}~\bibnamefont{Minkov}} \bibnamefont{and}
  \bibinfo{author}{\bibfnamefont{A.}~\bibnamefont{P\'alffy}},
  \bibinfo{journal}{Phys. Rev. Lett.} \textbf{\bibinfo{volume}{122}},
  \bibinfo{pages}{162502} (\bibinfo{year}{2019}).

\bibitem[{\citenamefont{Ruchowska et~al.}(2006)\citenamefont{Ruchowska,
  P\l{}\'ociennik, \ifmmode~\dot{Z}\else \.{Z}\fi{}ylicz, Mach, Kvasil, Algora,
  Amzal, B\"ack, Borge, Boutami et~al.}}]{PhysRevC.73.044326}
\bibinfo{author}{\bibfnamefont{E.}~\bibnamefont{Ruchowska}},
  \bibinfo{author}{\bibfnamefont{W.~A.} \bibnamefont{P\l{}\'ociennik}},
  \bibinfo{author}{\bibfnamefont{J.}~\bibnamefont{\ifmmode~\dot{Z}\else
  \.{Z}\fi{}ylicz}}, \bibinfo{author}{\bibfnamefont{H.}~\bibnamefont{Mach}},
  \bibinfo{author}{\bibfnamefont{J.}~\bibnamefont{Kvasil}},
  \bibinfo{author}{\bibfnamefont{A.}~\bibnamefont{Algora}},
  \bibinfo{author}{\bibfnamefont{N.}~\bibnamefont{Amzal}},
  \bibinfo{author}{\bibfnamefont{T.}~\bibnamefont{B\"ack}},
  \bibinfo{author}{\bibfnamefont{M.~G.} \bibnamefont{Borge}},
  \bibinfo{author}{\bibfnamefont{R.}~\bibnamefont{Boutami}},
  \bibnamefont{et~al.}, \bibinfo{journal}{Phys. Rev. C}
  \textbf{\bibinfo{volume}{73}}, \bibinfo{pages}{044326}
  (\bibinfo{year}{2006}).

\bibitem[{\citenamefont{Canty et~al.}(1977)\citenamefont{Canty, Connor, Dohan,
  and Pople}}]{Canty1977}
\bibinfo{author}{\bibfnamefont{M.~J.} \bibnamefont{Canty}},
  \bibinfo{author}{\bibfnamefont{R.~D.} \bibnamefont{Connor}},
  \bibinfo{author}{\bibfnamefont{D.~A.} \bibnamefont{Dohan}}, \bibnamefont{and}
  \bibinfo{author}{\bibfnamefont{B.}~\bibnamefont{Pople}},
  \bibinfo{journal}{Journal of Physics G: Nuclear Physics}
  \textbf{\bibinfo{volume}{3}}, \bibinfo{pages}{421} (\bibinfo{year}{1977}).

\bibitem[{\citenamefont{Guimar\~aes Filho and
  Helene}(2005)}]{PhysRevC.71.044303}
\bibinfo{author}{\bibfnamefont{Z.~O.} \bibnamefont{Guimar\~aes Filho}}
  \bibnamefont{and} \bibinfo{author}{\bibfnamefont{O.}~\bibnamefont{Helene}},
  \bibinfo{journal}{Phys. Rev. C} \textbf{\bibinfo{volume}{71}},
  \bibinfo{pages}{044303} (\bibinfo{year}{2005}).

\bibitem[{\citenamefont{Beck et~al.}(2007)\citenamefont{Beck, Becker,
  Beiersdorfer, Brown, Moody, Wilhelmy, Porter, Kilbourne, and
  Kelley}}]{PhysRevLett.98.142501}
\bibinfo{author}{\bibfnamefont{B.~R.} \bibnamefont{Beck}},
  \bibinfo{author}{\bibfnamefont{J.~A.} \bibnamefont{Becker}},
  \bibinfo{author}{\bibfnamefont{P.}~\bibnamefont{Beiersdorfer}},
  \bibinfo{author}{\bibfnamefont{G.~V.} \bibnamefont{Brown}},
  \bibinfo{author}{\bibfnamefont{K.~J.} \bibnamefont{Moody}},
  \bibinfo{author}{\bibfnamefont{J.~B.} \bibnamefont{Wilhelmy}},
  \bibinfo{author}{\bibfnamefont{F.~S.} \bibnamefont{Porter}},
  \bibinfo{author}{\bibfnamefont{C.~A.} \bibnamefont{Kilbourne}},
  \bibnamefont{and} \bibinfo{author}{\bibfnamefont{R.~L.}
  \bibnamefont{Kelley}}, \bibinfo{journal}{Phys. Rev. Lett.}
  \textbf{\bibinfo{volume}{98}}, \bibinfo{pages}{142501}
  (\bibinfo{year}{2007}).

\bibitem[{\citenamefont{Seiferle et~al.}(2019)\citenamefont{Seiferle, von~der
  Wense, Bilous, Amersdorffer, Lemell, Libisch, Stellmer, Schumm, Düllmann,
  Pálffy et~al.}}]{Seiferle2019}
\bibinfo{author}{\bibfnamefont{B.}~\bibnamefont{Seiferle}},
  \bibinfo{author}{\bibfnamefont{L.}~\bibnamefont{von~der Wense}},
  \bibinfo{author}{\bibfnamefont{P.~V.} \bibnamefont{Bilous}},
  \bibinfo{author}{\bibfnamefont{I.}~\bibnamefont{Amersdorffer}},
  \bibinfo{author}{\bibfnamefont{C.}~\bibnamefont{Lemell}},
  \bibinfo{author}{\bibfnamefont{F.}~\bibnamefont{Libisch}},
  \bibinfo{author}{\bibfnamefont{S.}~\bibnamefont{Stellmer}},
  \bibinfo{author}{\bibfnamefont{T.}~\bibnamefont{Schumm}},
  \bibinfo{author}{\bibfnamefont{C.~E.} \bibnamefont{Düllmann}},
  \bibinfo{author}{\bibfnamefont{A.}~\bibnamefont{Pálffy}},
  \bibnamefont{et~al.}, \bibinfo{journal}{Nature}
  \textbf{\bibinfo{volume}{573}}, \bibinfo{pages}{243} (\bibinfo{year}{2019}).

\bibitem[{\citenamefont{Kraemer et~al.}(2023)\citenamefont{Kraemer, Moens,
  Athanasakis-Kaklamanakis, Bara, Beeks, Chhetri, Chrysalidis, Claessens,
  Cocolios, Correia et~al.}}]{Kraemer2023}
\bibinfo{author}{\bibfnamefont{S.}~\bibnamefont{Kraemer}},
  \bibinfo{author}{\bibfnamefont{J.}~\bibnamefont{Moens}},
  \bibinfo{author}{\bibfnamefont{M.}~\bibnamefont{Athanasakis-Kaklamanakis}},
  \bibinfo{author}{\bibfnamefont{S.}~\bibnamefont{Bara}},
  \bibinfo{author}{\bibfnamefont{K.}~\bibnamefont{Beeks}},
  \bibinfo{author}{\bibfnamefont{P.}~\bibnamefont{Chhetri}},
  \bibinfo{author}{\bibfnamefont{K.}~\bibnamefont{Chrysalidis}},
  \bibinfo{author}{\bibfnamefont{A.}~\bibnamefont{Claessens}},
  \bibinfo{author}{\bibfnamefont{T.~E.} \bibnamefont{Cocolios}},
  \bibinfo{author}{\bibfnamefont{J.~G.~M.} \bibnamefont{Correia}},
  \bibnamefont{et~al.}, \bibinfo{journal}{Nature}
  \textbf{\bibinfo{volume}{617}}, \bibinfo{pages}{706} (\bibinfo{year}{2023}).

\bibitem[{\citenamefont{Campbell et~al.}(2011)\citenamefont{Campbell, Radnaev,
  and Kuzmich}}]{PhysRevLett.106.223001}
\bibinfo{author}{\bibfnamefont{C.~J.} \bibnamefont{Campbell}},
  \bibinfo{author}{\bibfnamefont{A.~G.} \bibnamefont{Radnaev}},
  \bibnamefont{and} \bibinfo{author}{\bibfnamefont{A.}~\bibnamefont{Kuzmich}},
  \bibinfo{journal}{Phys. Rev. Lett.} \textbf{\bibinfo{volume}{106}},
  \bibinfo{pages}{223001} (\bibinfo{year}{2011}).

\bibitem[{\citenamefont{{Gerstenkorn, S.}
  et~al.}(1974)\citenamefont{{Gerstenkorn, S.}, {Luc, P.}, {Verges, J.},
  {Englekemeir, D.W.}, {Gindler, J.E.}, and {Tomkins, F.S.}}}]{th229ground}
\bibinfo{author}{\bibnamefont{{Gerstenkorn, S.}}},
  \bibinfo{author}{\bibnamefont{{Luc, P.}}},
  \bibinfo{author}{\bibnamefont{{Verges, J.}}},
  \bibinfo{author}{\bibnamefont{{Englekemeir, D.W.}}},
  \bibinfo{author}{\bibnamefont{{Gindler, J.E.}}}, \bibnamefont{and}
  \bibinfo{author}{\bibnamefont{{Tomkins, F.S.}}}, \bibinfo{journal}{J. Phys.
  France} \textbf{\bibinfo{volume}{35}}, \bibinfo{pages}{483}
  (\bibinfo{year}{1974}).

\bibitem[{\citenamefont{Safronova et~al.}(2013)\citenamefont{Safronova,
  Safronova, Radnaev, Campbell, and Kuzmich}}]{PhysRevA.88.060501}
\bibinfo{author}{\bibfnamefont{M.~S.} \bibnamefont{Safronova}},
  \bibinfo{author}{\bibfnamefont{U.~I.} \bibnamefont{Safronova}},
  \bibinfo{author}{\bibfnamefont{A.~G.} \bibnamefont{Radnaev}},
  \bibinfo{author}{\bibfnamefont{C.~J.} \bibnamefont{Campbell}},
  \bibnamefont{and} \bibinfo{author}{\bibfnamefont{A.}~\bibnamefont{Kuzmich}},
  \bibinfo{journal}{Phys. Rev. A} \textbf{\bibinfo{volume}{88}},
  \bibinfo{pages}{060501} (\bibinfo{year}{2013}).

\bibitem[{\citenamefont{Thielking et~al.}(2018)\citenamefont{Thielking,
  Okhapkin, Głowacki, Meier, von~der Wense, Seiferle, Düllmann, Thirolf, and
  Peik}}]{Thielking2018}
\bibinfo{author}{\bibfnamefont{J.}~\bibnamefont{Thielking}},
  \bibinfo{author}{\bibfnamefont{M.~V.} \bibnamefont{Okhapkin}},
  \bibinfo{author}{\bibfnamefont{P.}~\bibnamefont{Głowacki}},
  \bibinfo{author}{\bibfnamefont{D.~M.} \bibnamefont{Meier}},
  \bibinfo{author}{\bibfnamefont{L.}~\bibnamefont{von~der Wense}},
  \bibinfo{author}{\bibfnamefont{B.}~\bibnamefont{Seiferle}},
  \bibinfo{author}{\bibfnamefont{C.~E.} \bibnamefont{Düllmann}},
  \bibinfo{author}{\bibfnamefont{P.~G.} \bibnamefont{Thirolf}},
  \bibnamefont{and} \bibinfo{author}{\bibfnamefont{E.}~\bibnamefont{Peik}},
  \bibinfo{journal}{Nature} \textbf{\bibinfo{volume}{556}},
  \bibinfo{pages}{321} (\bibinfo{year}{2018}).

\bibitem[{\citenamefont{Yoon et~al.}(2019)\citenamefont{Yoon, Jeon, Shin, Lee,
  Lee, Choi, Kim, Sung, and Nam}}]{Yoon:19}
\bibinfo{author}{\bibfnamefont{J.~W.} \bibnamefont{Yoon}},
  \bibinfo{author}{\bibfnamefont{C.}~\bibnamefont{Jeon}},
  \bibinfo{author}{\bibfnamefont{J.}~\bibnamefont{Shin}},
  \bibinfo{author}{\bibfnamefont{S.~K.} \bibnamefont{Lee}},
  \bibinfo{author}{\bibfnamefont{H.~W.} \bibnamefont{Lee}},
  \bibinfo{author}{\bibfnamefont{I.~W.} \bibnamefont{Choi}},
  \bibinfo{author}{\bibfnamefont{H.~T.} \bibnamefont{Kim}},
  \bibinfo{author}{\bibfnamefont{J.~H.} \bibnamefont{Sung}}, \bibnamefont{and}
  \bibinfo{author}{\bibfnamefont{C.~H.} \bibnamefont{Nam}},
  \bibinfo{journal}{Opt. Express} \textbf{\bibinfo{volume}{27}},
  \bibinfo{pages}{20412} (\bibinfo{year}{2019}).

\bibitem[{\citenamefont{Danson et~al.}(2004)\citenamefont{Danson, Brummitt,
  Clarke, Collier, Fell, Frackiewicz, Hancock, Hawkes, Hernandez-Gomez,
  Holligan et~al.}}]{Danson_2004}
\bibinfo{author}{\bibfnamefont{C.}~\bibnamefont{Danson}},
  \bibinfo{author}{\bibfnamefont{P.}~\bibnamefont{Brummitt}},
  \bibinfo{author}{\bibfnamefont{R.}~\bibnamefont{Clarke}},
  \bibinfo{author}{\bibfnamefont{J.}~\bibnamefont{Collier}},
  \bibinfo{author}{\bibfnamefont{B.}~\bibnamefont{Fell}},
  \bibinfo{author}{\bibfnamefont{A.}~\bibnamefont{Frackiewicz}},
  \bibinfo{author}{\bibfnamefont{S.}~\bibnamefont{Hancock}},
  \bibinfo{author}{\bibfnamefont{S.}~\bibnamefont{Hawkes}},
  \bibinfo{author}{\bibfnamefont{C.}~\bibnamefont{Hernandez-Gomez}},
  \bibinfo{author}{\bibfnamefont{P.}~\bibnamefont{Holligan}},
  \bibnamefont{et~al.}, \bibinfo{journal}{Nuclear Fusion}
  \textbf{\bibinfo{volume}{44}}, \bibinfo{pages}{S239} (\bibinfo{year}{2004}).

\bibitem[{\citenamefont{Waxer et~al.}(2005)\citenamefont{Waxer, Maywar, Kelly,
  Kessler, Kruschwitz, Loucks, McCrory, Meyerhofer, Morse, Stoeckl
  et~al.}}]{Waxer2005}
\bibinfo{author}{\bibfnamefont{L.}~\bibnamefont{Waxer}},
  \bibinfo{author}{\bibfnamefont{D.}~\bibnamefont{Maywar}},
  \bibinfo{author}{\bibfnamefont{J.}~\bibnamefont{Kelly}},
  \bibinfo{author}{\bibfnamefont{T.}~\bibnamefont{Kessler}},
  \bibinfo{author}{\bibfnamefont{B.}~\bibnamefont{Kruschwitz}},
  \bibinfo{author}{\bibfnamefont{S.}~\bibnamefont{Loucks}},
  \bibinfo{author}{\bibfnamefont{R.}~\bibnamefont{McCrory}},
  \bibinfo{author}{\bibfnamefont{D.}~\bibnamefont{Meyerhofer}},
  \bibinfo{author}{\bibfnamefont{S.}~\bibnamefont{Morse}},
  \bibinfo{author}{\bibfnamefont{C.}~\bibnamefont{Stoeckl}},
  \bibnamefont{et~al.}, \bibinfo{journal}{Opt. Photon. News}
  \textbf{\bibinfo{volume}{16}}, \bibinfo{pages}{30} (\bibinfo{year}{2005}).

\bibitem[{\citenamefont{Tiwari et~al.}(2019)\citenamefont{Tiwari, Gaul,
  Martinez, Dyer, Gordon, Spinks, Toncian, Bowers, Jiao, Kupfer
  et~al.}}]{Tiwari2019}
\bibinfo{author}{\bibfnamefont{G.}~\bibnamefont{Tiwari}},
  \bibinfo{author}{\bibfnamefont{E.}~\bibnamefont{Gaul}},
  \bibinfo{author}{\bibfnamefont{M.}~\bibnamefont{Martinez}},
  \bibinfo{author}{\bibfnamefont{G.}~\bibnamefont{Dyer}},
  \bibinfo{author}{\bibfnamefont{J.}~\bibnamefont{Gordon}},
  \bibinfo{author}{\bibfnamefont{M.}~\bibnamefont{Spinks}},
  \bibinfo{author}{\bibfnamefont{T.}~\bibnamefont{Toncian}},
  \bibinfo{author}{\bibfnamefont{B.}~\bibnamefont{Bowers}},
  \bibinfo{author}{\bibfnamefont{X.}~\bibnamefont{Jiao}},
  \bibinfo{author}{\bibfnamefont{R.}~\bibnamefont{Kupfer}},
  \bibnamefont{et~al.}, \bibinfo{journal}{Opt. Lett.}
  \textbf{\bibinfo{volume}{44}}, \bibinfo{pages}{2764} (\bibinfo{year}{2019}).

\bibitem[{\citenamefont{Alder et~al.}(1956)\citenamefont{Alder, Bohr, Huus,
  Mottelson, and Winther}}]{RevModPhys.28.432}
\bibinfo{author}{\bibfnamefont{K.}~\bibnamefont{Alder}},
  \bibinfo{author}{\bibfnamefont{A.}~\bibnamefont{Bohr}},
  \bibinfo{author}{\bibfnamefont{T.}~\bibnamefont{Huus}},
  \bibinfo{author}{\bibfnamefont{B.}~\bibnamefont{Mottelson}},
  \bibnamefont{and} \bibinfo{author}{\bibfnamefont{A.}~\bibnamefont{Winther}},
  \bibinfo{journal}{Rev. Mod. Phys.} \textbf{\bibinfo{volume}{28}},
  \bibinfo{pages}{432} (\bibinfo{year}{1956}).

\bibitem[{\citenamefont{Alder and Winther}(1954)}]{PhysRev.96.237}
\bibinfo{author}{\bibfnamefont{K.}~\bibnamefont{Alder}} \bibnamefont{and}
  \bibinfo{author}{\bibfnamefont{A.}~\bibnamefont{Winther}},
  \bibinfo{journal}{Phys. Rev.} \textbf{\bibinfo{volume}{96}},
  \bibinfo{pages}{237} (\bibinfo{year}{1954}).

\bibitem[{\citenamefont{Biedenharn et~al.}(1955)\citenamefont{Biedenharn,
  McHale, and Thaler}}]{Biedenharn:1955zza}
\bibinfo{author}{\bibfnamefont{L.~C.} \bibnamefont{Biedenharn}},
  \bibinfo{author}{\bibfnamefont{J.~L.} \bibnamefont{McHale}},
  \bibnamefont{and} \bibinfo{author}{\bibfnamefont{R.~M.}
  \bibnamefont{Thaler}}, \bibinfo{journal}{Phys. Rev.}
  \textbf{\bibinfo{volume}{100}}, \bibinfo{pages}{376} (\bibinfo{year}{1955}).

\bibitem[{\citenamefont{Biedenharn et~al.}(1956)\citenamefont{Biedenharn,
  Goldstein, McHale, and Thaler}}]{PhysRev.101.662}
\bibinfo{author}{\bibfnamefont{L.~C.} \bibnamefont{Biedenharn}},
  \bibinfo{author}{\bibfnamefont{M.}~\bibnamefont{Goldstein}},
  \bibinfo{author}{\bibfnamefont{J.~L.} \bibnamefont{McHale}},
  \bibnamefont{and} \bibinfo{author}{\bibfnamefont{R.~M.}
  \bibnamefont{Thaler}}, \bibinfo{journal}{Phys. Rev.}
  \textbf{\bibinfo{volume}{101}}, \bibinfo{pages}{662} (\bibinfo{year}{1956}).

\bibitem[{\citenamefont{Eisenberg and Greiner}(1976)}]{Eisenberg1976}
\bibinfo{author}{\bibfnamefont{J.}~\bibnamefont{Eisenberg}} \bibnamefont{and}
  \bibinfo{author}{\bibfnamefont{W.}~\bibnamefont{Greiner}},
  \emph{\bibinfo{title}{Nuclear Theory: Excitation mechanisms of the nucleus.
  Vol.2}}, Nuclear theory (\bibinfo{publisher}{North-Holland},
  \bibinfo{year}{1976}), ISBN \bibinfo{isbn}{9780720404838}.

\bibitem[{\citenamefont{Bohr and Mottelson}(1998)}]{Bohr1998}
\bibinfo{author}{\bibfnamefont{A.}~\bibnamefont{Bohr}} \bibnamefont{and}
  \bibinfo{author}{\bibfnamefont{B.~R.} \bibnamefont{Mottelson}},
  \emph{\bibinfo{title}{Nuclear structure.}} (\bibinfo{publisher}{World
  Scientific}, \bibinfo{year}{1998}), ISBN \bibinfo{isbn}{9810231970}.

\bibitem[{\citenamefont{Ring and Schuck}(2004)}]{ring2004nuclear}
\bibinfo{author}{\bibfnamefont{P.}~\bibnamefont{Ring}} \bibnamefont{and}
  \bibinfo{author}{\bibfnamefont{P.}~\bibnamefont{Schuck}},
  \emph{\bibinfo{title}{The Nuclear Many-Body Problem}}, Physics and astronomy
  online library (\bibinfo{publisher}{Springer}, \bibinfo{year}{2004}), ISBN
  \bibinfo{isbn}{9783540212065}.

\end{thebibliography}

\end {document}